# Transient Stability of GFL Converters Subjected to Switching of Droop-Controlled GFM Converters

Bingfang Li, *Member, IEEE*, Songhao Yang, *Senior Member, IEEE*,
Pu Cheng, *Graduate Student Member, IEEE*, Zhiguo Hao, *Senior Member*, *IEEE*

*Abstract*—**Integrating grid-forming converters (GFMCs) into grid-following converter (GFLC)-dominated power systems enhances the grid strength, but GFMCs' current-limiting characteristic triggers dynamic switching between constant voltage control (CVC) and current limit control (CLC). This switching feature poses critical transient stability risks to GFLCs, requiring urgent investigation. This paper first develops a mathematical model for this switched system. Then, it derives switching conditions for droop-controlled GFMCs, which are separately GFMC angle-dependent and GFLC angle-dependent. On this basis, the stability boundaries of GFLC within each subsystem are analyzed, and the impact of GFMC switching arising from GFLC angle oscillation is investigated. The findings reveal that the switched system's stability boundary coincides with that of the CLC subsystem. To enhance GFLC's transient stability and ensure GFMC converges to the CVC mode, this paper introduces a virtual fixed d-axis control (VFDC) strategy. Compared with existing methods, this method achieves decoupling and self-stabilization using only local state variables from individual converters. The conclusions are validated through simulations and Controller Hardware-in-the-Loop tests.**

*Index Terms*—**Grid forming, Grid following, Phase locked loops, Stability analysis, Power system control, Current limiter**

## NOMENCLATURE

$U_{c1}\angle\delta_0$ Voltage phasor of the GFMC.

$U_{c2}\angle\theta_0$ Voltage phasor of the GFLC.

$U_g\angle 0°$ Voltage phasor of the infinite bus.

$I_{c1}\angle\varphi_{c1}$ Current phasor of GFMC.

$I_{c2}\angle\varphi_{c2}$ Current phasor of GFLC.

$Y_{c1}$, $L_{c1}$ Admittance, inductance between GFMC terminal and PCC.

$Y_{c2}$, $L_{c2}$ Admittance, inductance between GFLC terminal and PCC.

$Y_g$, $L_g$ Admittance, inductance between PCC and the infinite bus.

$\alpha$ Defined as $Y_{c1}/(Y_{c1}+Y_g)$.

$Y_{1g}$ Defined as $Y_{1g}=Y_{c1}Y_g/(Y_{c1}+Y_g)$.

$L_v$ Defined as $L_{c2}+(1-\alpha)\cdot L_g$.

$\beta_1$ Defined as $1/|Y_g|+1/|Y_{c1}|$.

$\gamma$ Defined as $1/|Y_g|$.

$d_1$-$q_1$, PSL rotating frame.

$d_2$-$q_2$ PLL rotating frame.

x-y Synchronous rotating frame.

$\omega_{c1}$, $\omega_{c2}$, $\omega_g$ Angular frequency of PSL, PLL, and infinite bus.

$\varpi$ Defined as $\omega_{c2}$- $\omega_g$.

$\delta$ PSL angle relative to the infinite bus.

$\theta$ PLL angle relative to the infinite bus.

$\delta_{ins}$, $\theta_{ins}$, Instantaneous angle of PSL, PLL.

$\varphi_{c1}$, $\eta_1$ Angles of GFMC current phasor relative to the infinite bus, to the $d_1$-axis.

$\varphi_{c2}$, $\eta_2$ Angles of GFLC current phasor relative to the infinite bus, to the $d_2$-axis.

$\varphi_{c1*}$,$\varphi_{12*}$ Current angles applying VFDC.

$\delta_k^{L}$, $\delta_k^{R}$ GFMC switching boundaries in period *k*.

$\delta_p$ Fixed PSL angle during PLL dynamics.

$\theta_s^{V}$, $\theta_s^{L}$ SEP angle of PLL in CVC, CLC subsystem

$\theta_{max}^{V}$, $\theta_{min}^{V}$ Maximum, minimum stability boundary angles of the PLL in the CVC subsystem.

$\theta_{max}^{L}$, $\theta_{min}^{L}$ Maximum, minimum stability boundary angles of the PLL in the CLC subsystem.

$\dot{i}_{c1}^{dref}$, $\dot{i}_{c1}^{qref}$ GFMC current references.

$\dot{i}_{c2}^{dref}$, $\dot{i}_{c2}^{qref}$ GFLC current references.

$\dot{i}_{c1*}^{dref}$, $\dot{i}_{c1*}^{qref}$ GFMC current references applying VFDC.

$\dot{i}_{c2*}^{dref}$, $\dot{i}_{c2*}^{qref}$ GFLC current references applying VFDC.

$\dot{i}_{c1}^{d_1}$, $\dot{i}_{c1}^{q_1}$ GFMC $d_1$-axis, $q_1$-axis currents.

$\dot{i}_{c2}^{d_2}$, $\dot{i}_{c2}^{q_2}$ GFLC $d_1$-axis, $q_1$-axis currents.

$\dot{i}_{c1}^{d_2}$, $\dot{i}_{c1}^{q_2}$ $d_2$ axis, $q_2$ axis components of $I_{c1}\angle\varphi_{c1}$.

$\dot{i}_{c2ij}^{d_2}$ Current coupling term between GFLCs

$I_{c1}^{max}$ Maximum GFMC current limit.

$I_{c2}^{ref}$ GFLC current reference amplitude.

$u_{c2}^{q_2}$ $q_2$-axis component of $U_{c2}\angle\theta_0$.

$u_{c2(V)}^{q_2}$, $u_{c2(L)}^{q_2}$ $q_2$-axis component of $U_{c2}\angle\theta_0$ in CVC, CLC.

$k_{2p}$, $k_{2I}$ Proportional and integral gain of PLL.

$P_{c1}^{ref}$, $P_{c1}$ Reference, actual active power of GFMC.

$P_{c1}^{CVC}$,$P_{c1}^{CLC}$ Active power of GFMC in CVC, CLC.

$P_{c1*}^{CLC}$ Active power of GFMC in CLC with VFDC

$P_{Ec2}^{V}$, $P_{Ec2}^{L}$ GFLC equivalent electromagnetic power in the CVC, CLC subsystem.

$P_{Mc2}^{V}$, $P_{Mc2}^{L}$ GFLC equivalent mechanical power in the CVC, CLC subsystem.

$T_{c2}^{V}$, $T_{c2}^{L}$ GFLC equivalent inertia time constant in the CVC, CLC subsystem.

$D_{c2}^{V}$, $D_{c2}^{L}$ GFLC equivalent damping coefficient in the CVC, CLC subsystem.

$D_{c2*}^{L}$ GFLC equivalent second-order damping coefficient in the CLC subsystem.

$\Gamma_V$, $\Gamma_L$, $\Gamma_{L1}$, $\Gamma_{L2}$ Vector trajectories.

$V$ Transient energy of PLL.

$V_{max}^{V}$, $V_{max}^{L}$ Critical energy of PLL in CVC, CLC.

$\Delta V_o$ Periodic energy increment of PLL.

$\Delta V_p^{E}$, $\Delta V_p^{F}$ Potential energy discontinuities at *E* and *F*.

$\Delta V_D$ Periodic change in *V* due to damping.

$Z_{virtual}$ Virtual impedance of GFMC.

# I. INTRODUCTION

THE integration of renewable energy sources into modern power grid is rapidly increasing, primarily using grid following converters (GFLCs) for grid connections[1],[2]. Extensive studies indicate that systems with high penetration of renewable energy and long-distance power transmission are significantly more susceptible to wide-band oscillations and frequency/voltage accidents [1],[3]. Moreover, major blackouts like the recent "4.28" incident in Spain[4] starkly illustrate that weak disturbance rejection of GFLCs is a key trigger for widespread outages. Consequently, grid-forming (GFM) devices are suggested to be deployed within renewable energy stations to bolster the strength and inertia of the sending-end system[5],[6].

In grid-connected scenarios, two main types of grid-forming converters (GFMCs) are applied: those based on the power synchronization loop (PSL)[7] and those utilizing virtual synchronous generator (VSG) control[8],[9]. VSG-GFMCs emulate synchronous generators (SGs) to provide inertia and reactivate power support; however, they are susceptible to transient angle instability. In contrast, PSL-based GFMCs (also known as droop-controlled GFMCs) are inherently stable as a first-order control system if a stable equilibrium point (SEP) exists. Thus, despite exhibiting minor steady-state errors, the inherent stability and simpler control architecture of PSL-GFMCs make them a highly promising technology.

However, GFMCs can switch from constant voltage control (CVC) mode to current limit control (CLC) to avoid overcurrent during transients[11]. This switching behavior transforms the GFLC-dominant system into a switched system, posing new challenges to GFLCs' transient stability. Current research on the transient stability of this co-located system primarily focuses on GFLCs operating in parallel with SGs or VSG-GFMCs. For instance, ref. [12] investigates the impact of GFLC's fault current angle on VSG-GFMC stability; however, it does not address the transient stability of GFLC's phase-locked loop (PLL). Refs. [13] and [14] examines the stability issues of GFMC and GFLC. Specifically, ref. [13] focuses on the impact of voltage drops at the point of common coupling (PCC), caused by GFMC angle swings, on the PLL stability. Ref. [14] investigates the dynamic interaction between GFMCs and GFLCs using the phase trajectory method. However, these studies overlook the impact of GFMC's current limiting control on the transient stability of nearby GFLCs.

In such systems, in addition to the transient instability issues faced by PLLs, the system's potential failure to ultimately stabilize in CVC mode presents another significant concern. Therefore, measures to enhance system stability need to address two key aspects: PLLs' transient stability and GFMC stabilizing at the desired SEP.

To enhance PLL stability in multi-GFLC systems, refs [15] and [16] have proposed two different current distribution or injection methods. However, these approaches require comprehensive global information, which increases the complexity of implementation. Regarding the issue that GFMC stabilizes at the CLC rather than the CVC after fault clearance, ref. [17] proposes an anti-windup control strategy. But ref. [18] further demonstrates that, even with such control, GFMC stabilization in CVC mode hinges on the PSL angle at fault clearance. It then proposes a method to calculate a GFMC current saturation angle that ensures its SEP is precisely within the CVC region. However, this approach introduces significant complexity, as its calculation requires system-wide information and continuous updates based on the GFLC's operating state. For stability enhancement control in GFLC and GFMC co-located systems, the current angle coordination of converters proposed in [12] requires real-time communication and is mainly focused on improving VSG-GFMC stability. The additional power control for GFMC proposed in [13] does not aim to address its stabilization at the correct SEP. Therefore, an approach that concurrently addresses both the transient stability of GFLCs and desired equilibrium point restoration for droop-controlled GFMCs warrants further investigation.

In summary, this paper investigates the impact of droop-controlled GFMC switching on the transient synchronization stability of GFLC and further proposes a stability enhancement control strategy. The main contributions are as follows:

1) A math model for the switched system is established, where PLL dynamics are considered predominant, and PSL dynamics define the switching conditions. Two switching conditions for droop-controlled GFMC are identified: "GFMC angle-dependent Switching" and "GFLC angle-dependent Switching". This enables the analysis of the interactive coupling effects between the PLL and PSL.
2) The mechanism by which GFMC switching affects GFLC transient stability is elucidated. The study involves a comparison of PLL stability boundaries, encompassing the GFLC stability characteristics within the CVC and CLC subsystems and the influence of inter-subsystem switching on GFLC transient stability. This reveals that the transient stability of GFLC is constrained by the stability boundary of the CLC subsystem.
3) The Virtual Fixed d-axis Control (VFDC) method was proposed. This method enhances the transient stability of multiple GFLCs and ensures GFMC adaptive stabilization in CVC mode, utilizing only local state feedback.

The remainder of this paper is organized as follows. Section II details the mathematical model of the switched system. Section III then presents a theoretical analysis covering SEPs, the stability of each subsystem, and the impact of the switching process. Section IV introduces the proposed control strategy and provides its theoretical validations. Section V presents the simulation and experimental results. A thorough discussion is offered in Section VI, and Section VII concludes the paper.

# II. SWITCHED SYSTEM MODELING

In the co-located system shown in Fig. 1, the PLL's dynamic behavior is profoundly influenced by external switching events, namely the transitions of the GFMC between CVC and CLC modes. This section develops a model for the switched system in which PLL dynamics are considered the dominant factor governing system stability. The dynamics of the inherently stable droop-controlled GFMC define the switching conditions for the switched system. Consequently, the system operates within its CVC subsystem when the GFMC is in CVC mode, and within its CLC subsystem during CLC mode operation.

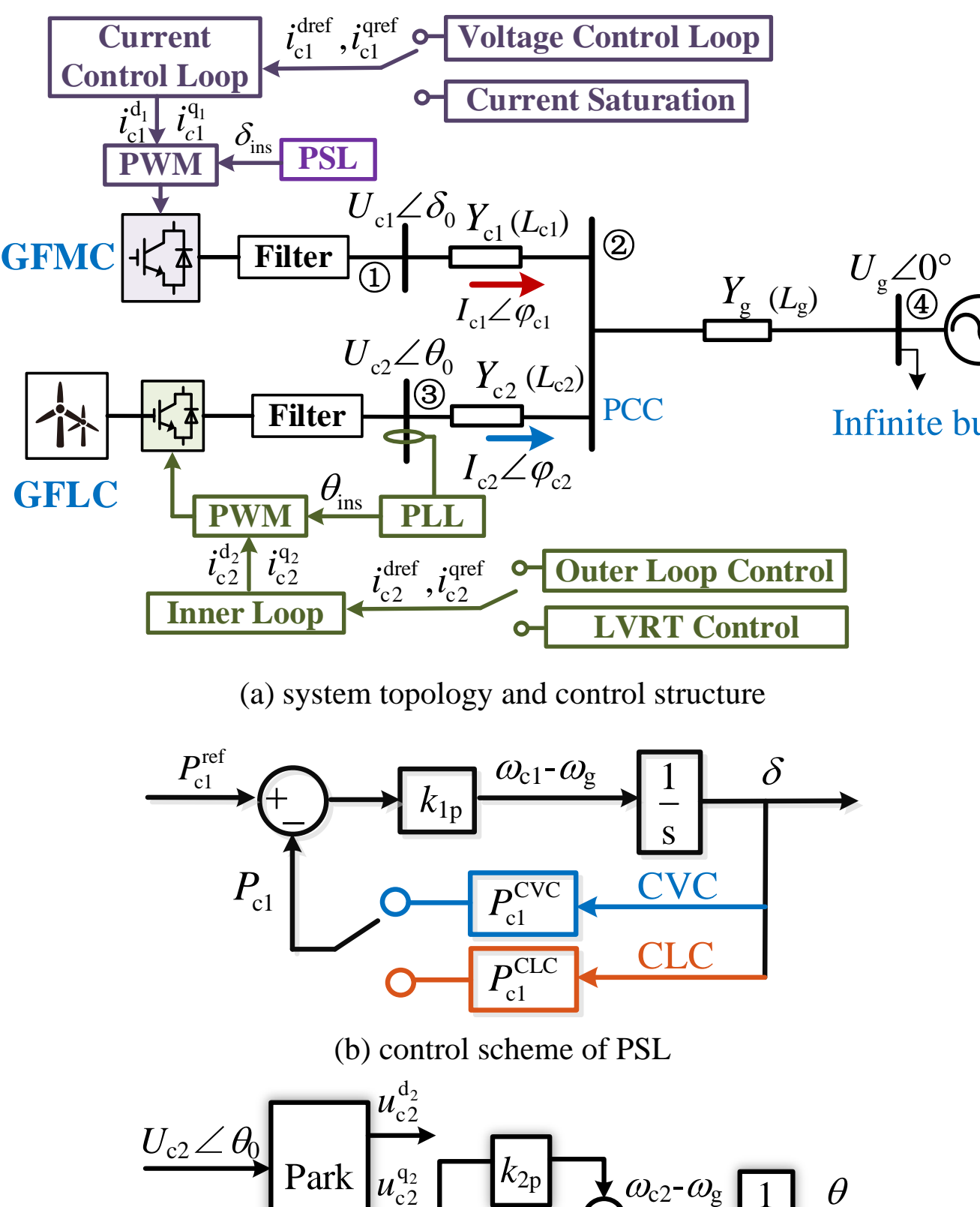

(a) system topology and control structure

(b) control scheme of PSL

(c) control scheme of PLL

**Fig. 1.** Topology and control structure of the co-located system

## *A. Co-located System Overview*

Fig. 1 depicts a topology of a GFLC-dominated renewable energy transmission system that incorporates droop-controlled GFMCs. These converters are aggregated at the same Point of Common Coupling (PCC) and connect to the receiving-end grid via AC transmission lines. $U_{c1}\angle\delta_0$, $U_{c2}\angle\theta_0$, $U_g\angle 0°$ denote the voltage phasors of the GFMC terminal, the GFLC terminal, and the infinite bus. $I_{c1}\angle\varphi_{c1}$, $I_{c2}\angle\varphi_{c2}$, and $I_g\angle\varphi_g$ are the corresponding branch currents, while $Y_{c1}$, $Y_{c2}$, and $Y_g$ are the branch admittances. The branch inductances are denoted as $L_{c1}$, $L_{c2}$, and $L_g$. Notably, for the centralized GFMCs, a master converter can generate control signals that the slave converters then follow[19]. This strategy mitigates potential control conflicts and the effects of circulating currents, allowing these multiple GFMCs to be effectively represented as a single-machine equivalent model for analytical purposes. Furthermore, to elucidate the fundamental mechanism by which GFMC switching impacts GFLC stability, GFLCs are also represented by aggregated models in the theoretical analysis, as shown in Fig. 1(a). However, for the subsequent control design and experimental validation stages, multi-GFLC models are employed to ensure the reliability of the research conclusions.

In Fig. 2(a), the $d_1$-$q_1$, $d_2$-$q_2$, and x-y coordinate systems are the PSL, PLL, and synchronous rotating frames, rotating at angular speeds $\omega_{c1}$, $\omega_{c2}$, and $\omega_g$, respectively. The angles $\delta$ and $\theta$ show how $d_1$ and $d_2$ lead the $x$-axis, corresponding to the output angles of the PSL and PLL. $\eta_1$ and $\eta_2$ are defined as angles by which $I_{c1}\angle\varphi_{c1}$ (in CLC mode) and $I_{c2}\angle\varphi_{c2}$ lead the $d_1$ and $d_2$ axes, respectively.

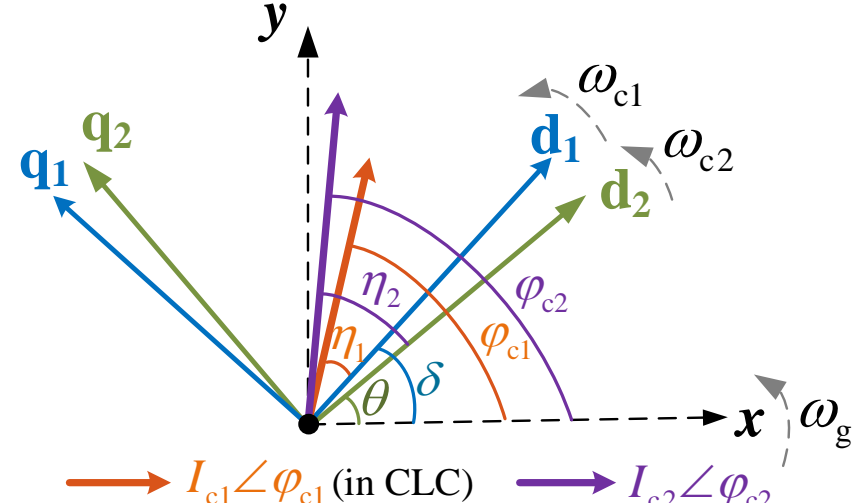

**Fig. 2.** Relation of reference axes and system vectors

## *B. Dynamic Model of the Switched System*

To facilitate analysis and focus on the primary issue of stability determination, the following reasonable assumptions are made:

1) In co-located systems, the GFMC is expected to provide a stable voltage and angle reference to facilitate robust phase locking by the GFLC [6]. Therefore, with proper parameter design, the bandwidth of the GFMC's PSL should be significantly smaller than that of the PLL. Concurrently, the voltage and current control loops are designed to be significantly faster than the PLL dynamics. Otherwise, undesirable dynamic coupling and small-signal stability issues would arise [19],[21],[22]. Consequently, when analyzing the GFLC's PLL dynamic response, the GFMC's angle $\delta$ can be considered quasi-constant, and its voltage and current control loops can be assumed to have converged [21],[22].

2) In typical weak sending-end systems, the electrical coupling between the GFMC and the PCC is significantly stronger than that between the PCC and the receiving-end grid. Furthermore, GFMCs in weak sending-ends are expected to provide substantial short-circuit current support [5],[13], which necessitates a small virtual and physical impedance between GFMC and PCC. Thus, we assume $Y_{c1}>>Y_g$.

3) Line resistances are neglected in comparison to line reactance [7],[8],[23],[24].

The PLL control scheme is shown in Fig. 1(c), and its control equations can be written as:

$$\begin{cases}\dfrac{\mathrm{d}\theta}{\mathrm{d}t}=\omega_{c2}-\omega_g=\varpi\\[2mm] \dfrac{\mathrm{d}\varpi}{\mathrm{d}t}=k_{2I}u_{c2}^{q_2}+k_{2p}\dfrac{\mathrm{d}u_{c2}^{q_2}}{\mathrm{d}t}\end{cases},\qquad(1)$$

where $k_{2p}$ and $k_{2I}$ are the proportional and integral coefficients of the PLL; $u_{c2}^{q_2}$ represents the component of $U_{c2}\angle\theta_0$ along the $q_2$-axis in the $d_2$-$q_2$ coordinate system, which is influenced by the GFMC's switching mode. If the GFMC current reference from the voltage control loop is below its upper limit, $U_{c1}\angle\delta_0$ aligns with the $d_1$-axis, with the $d_1$-axis component controlled to $U_{c1}$ and the $q_1$-axis component set to zero. This is known as CVC. If the current reference reaches its maximum, $I_{c1}^{max}$, the converter switches to CLC mode. This paper adopts a priority-based current limiter, as detailed in [10],[18]. During CLC operation, the saturation current angle $\eta_1$ (relative to the $d_1$-axis) is a configurable parameter, allowing for flexible prioritization strategies. The active power of the GFMC in CVC and CLC modes is denoted as $P_{\mathrm{c1}}^{\mathrm{CVC}}$ and $P_{\mathrm{c1}}^{\mathrm{CLC}}$, respectively, with their expressions given as:

$$\begin{cases} P_{\mathrm{c1}}^{\mathrm{CVC}} = U_{\mathrm{c1}} U_{\mathrm{g}} \left| Y_{\mathrm{1g}} \right| \sin\delta - \alpha U_{\mathrm{c1}} I_{\mathrm{c2}} \cos\left(\delta - \varphi_2\right) \\ P_{\mathrm{c1}}^{\mathrm{CLC}} = U_{\mathrm{g}} I_{\mathrm{c1}}^{\max} \cos\varphi_{\mathrm{c1}} - I_{\mathrm{c1}}^{\max} I_{\mathrm{c2}} / |Y_{\mathrm{g}}| \sin\left(\varphi_{\mathrm{c2}} - \varphi_{\mathrm{c1}}\right) \end{cases}, \tag{2}$$

where $\alpha=Y_{\mathrm{c1}}/(Y_{\mathrm{c1}}+Y_{\mathrm{g}})$, $Y_{\mathrm{1g}}=Y_{\mathrm{c1}}Y_{\mathrm{g}}/(Y_{\mathrm{c1}}+Y_{\mathrm{g}})$. Accordingly, within the CVC subsystem, $u_{\mathrm{c2}}^{\mathrm{q_2}}$ and its time derivative (denoted as $u_{\mathrm{c2(V)}}^{\mathrm{q_2}}$ and $\mathrm{d}u_{\mathrm{c2(V)}}^{\mathrm{q_2}}/\mathrm{d}t$) are derived as:

$$\begin{cases} u_{\mathrm{c2(V)}}^{\mathrm{q_2}} = \omega_{\mathrm{c2}} L_{\mathrm{v}} i_{\mathrm{c2}}^{\mathrm{d_2}} - \alpha U_{\mathrm{c1}} \sin\left(\theta - \delta\right) - (1-\alpha) U_{\mathrm{g}} \sin\theta \\ \dfrac{\mathrm{d}u_{\mathrm{c2(V)}}^{\mathrm{q_2}}}{\mathrm{d}t} = -\varpi\left(1-\alpha\right) U_{\mathrm{g}} \cos\theta - \varpi\alpha U_{\mathrm{c1}} \cos\left(\theta-\delta\right) + \dfrac{\mathrm{d}\varpi}{\mathrm{d}t}\left(1-\alpha\right) L_{\mathrm{v}} i_{\mathrm{c2}}^{\mathrm{d_2}} \end{cases}, \tag{3}$$

where $i_{\mathrm{c2}}^{\mathrm{d_2}}$ represents the active current of the GFLC; $L_{\mathrm{v}}=L_{\mathrm{c2}}+(1-\alpha)\cdot L_{\mathrm{g}}$. Similarly, within the CLC subsystem, $u_{\mathrm{c2}}^{\mathrm{q_2}}$ and its time derivative (denoted as $u_{\mathrm{c2(L)}}^{\mathrm{q_2}}$ and $\mathrm{d}u_{\mathrm{c2(L)}}^{\mathrm{q_2}}/\mathrm{d}t$) are derived as:

$$\begin{cases} u_{\mathrm{c2(L)}}^{\mathrm{q_2}} = -U_{\mathrm{g}} \sin\theta + \omega_{\mathrm{c2}} L_{\mathrm{g}} i_{\mathrm{c1}}^{\mathrm{d_2}} + \omega_{\mathrm{c2}}\left(L_{\mathrm{g}} + L_{\mathrm{c2}}\right) i_{\mathrm{c2}}^{\mathrm{d_2}} \\ \dfrac{\mathrm{d}u_{\mathrm{c2(L)}}^{\mathrm{q_2}}}{\mathrm{d}t} = -\varpi U_{\mathrm{g}} \cos\theta + \dfrac{\mathrm{d}\varpi}{\mathrm{d}t} L_{\mathrm{g}} i_{\mathrm{c1}}^{\mathrm{d_2}} + \varpi\left(\varpi + \omega_{\mathrm{g}}\right) L_{\mathrm{g}} i_{\mathrm{c1}}^{\mathrm{q_2}} + \dfrac{\mathrm{d}\varpi}{\mathrm{d}t}\left(L_{\mathrm{g}} + L_{\mathrm{c2}}\right) i_{\mathrm{c2}}^{\mathrm{d_2}} \end{cases}, \tag{4}$$

where $i_{\mathrm{c1}}^{\mathrm{d_2}}$ and $i_{\mathrm{c1}}^{\mathrm{q_2}}$ are the projections of the GFMC current onto the $d_2$-axis and $q_2$-axis, respectively. They are expressed as:

$$i_{\mathrm{c1}}^{\mathrm{d_2}} = I_{\mathrm{c1}} \cos\left(\varphi_{\mathrm{c1}} - \theta\right), \quad i_{\mathrm{c1}}^{\mathrm{q_2}} = I_{\mathrm{c1}} \sin\left(\varphi_{\mathrm{c1}} - \theta\right). \tag{5}$$

Substituting (3) and (4) into (1), the dynamic equations of the switched system are obtained as follows:

$$\frac{\mathrm{d}^2\theta}{\mathrm{d}t^2} = \frac{\mathrm{d}\varpi}{\mathrm{d}t} = \begin{cases} \dfrac{1}{T_{\mathrm{c2}}^{\mathrm{V}}}\left(P_{\mathrm{Mc2}}^{\mathrm{V}} - P_{\mathrm{Ec2}}^{\mathrm{V}} - D_{\mathrm{c2}}^{\mathrm{V}}\varpi\right), \text{ in CVC subsystem} \\ \dfrac{1}{T_{\mathrm{c2}}^{\mathrm{L}}}\left(P_{\mathrm{Mc2}}^{\mathrm{L}} - P_{\mathrm{Ec2}}^{\mathrm{L}} - D_{\mathrm{c2}}^{\mathrm{L}}\varpi - D_{\mathrm{c2*}}^{\mathrm{L}}\varpi^2\right), \text{ in CLC subsystem} \end{cases}, \tag{6}$$

where $P_{\mathrm{Ec2}}^{\mathrm{V}}$, $P_{\mathrm{Mc2}}^{\mathrm{V}}$, $T_{\mathrm{c2}}^{\mathrm{V}}$, and $D_{\mathrm{c2}}^{\mathrm{V}}$ are the terms in the CVC subsystem; $P_{\mathrm{Ec2}}^{\mathrm{L}}$, $P_{\mathrm{Mc2}}^{\mathrm{L}}$, $T_{\mathrm{c2}}^{\mathrm{L}}$, $D_{\mathrm{c2}}^{\mathrm{L}}$, and $D_{\mathrm{c2*}}^{\mathrm{L}}$ are the terms in the CLC subsystem. For detailed expressions for these variables, please refer to Appendix A.

*C. Switching Conditions*

The most fundamental condition for the GFMC to switch from CVC mode to CLC mode is whether its current amplitude $I_{\mathrm{c1}}$ equals the maximum allowable value $I_{\mathrm{c1}}^{\max}$. After fault clearance, this switching condition is:

$$I_{\mathrm{c1}} = \left| Y_{\mathrm{c1}} \left[ U_{\mathrm{c1}}\angle\delta_0 - \left( I_{\mathrm{c1}}\angle\left(\delta + \eta_1\right) + I_{\mathrm{c2}}\angle\theta \right) / Y_{\mathrm{g}} - U_{\mathrm{g}}\angle 0^\circ \right] \right| = I_{\mathrm{c1}}^{\max} \tag{7}$$

To facilitate the analysis of GFMC switching dynamics, the condition in (7) needs to be expressed using state variables. As (7) indicates, this switching condition is influenced by both $\delta$ and $\theta$. Consequently, the system's switching conditions will be derived based on their dependence on the GFMC angle and the GFLC angle, respectively.

**1) GFMC angle-dependent Switching Conditions**

In the CVC subsystem, the PSL angle is the phase angle of the voltage constructed by the GFMC, so $\delta=\delta_0$, where $\delta_0$ is the voltage angle at node ① in Fig. 1(a). Substituting this condition into (7), the range of $\delta$ that keeps the CVC subsystem active is:

$$\delta_k^{\mathrm{L}} \le \delta \le \delta_k^{\mathrm{R}}, \begin{cases} \delta_k^{\mathrm{R}} = \arccos\left(d\right) + \pi/2 - \lambda + 2k\pi, k \in Z \\ \delta_k^{\mathrm{L}} = -\arccos\left(d\right) + \pi/2 - \lambda + 2k\pi, k \in Z \end{cases}$$

$$\text{with} \begin{cases} \lambda = \arctan\left[\left(|Y_{\mathrm{g}}| U_{\mathrm{g}} - I_{\mathrm{c2}} \sin\varphi_{\mathrm{c2}}\right) / \left(I_{\mathrm{c2}} \cos\varphi_{\mathrm{c2}}\right)\right] \\ d = \dfrac{U_{\mathrm{c1}}^{\ 2} + U_{\mathrm{g}}^{\ 2} + \left(I_{c2}/|Y_g|\right)^2 - \left(I_{\mathrm{c1}}^{\max}/\left|Y_{\mathrm{1g}}\right|\right)^2 - \left(2U_{\mathrm{g}} I_{\mathrm{c2}}/|Y_{\mathrm{g}}|\right)\sin\varphi_{\mathrm{c2}}}{\sqrt{\left(2U_{\mathrm{c1}} I_{\mathrm{c2}}/|Y_{\mathrm{g}}|\right)^2 + \left(2U_{\mathrm{g}} U_{\mathrm{c1}}\right)^2 - \left(8U_{\mathrm{g}} U_{\mathrm{c1}}^{\ 2} I_{\mathrm{c2}}/|Y_{\mathrm{g}}|\right)\sin\varphi_{\mathrm{c2}}}} \end{cases}. \tag{8}$$

In (8), $\delta=\delta_k^{\mathrm{L}}$ and $\delta=\delta_k^{\mathrm{R}}$ define the switching boundaries for the system in period $k$, which are also the GFMC angle-dependent switching conditions. This results in an alternating pattern of CVC and CLC intervals in the $P_{\mathrm{c1}}$-$\delta$ plane, as shown in Fig. 3, where $P_{\mathrm{c1}}$ represents the active power of the GFMC. After fault clearance, if $\delta_k^{\mathrm{L}}<\delta<\delta_k^{\mathrm{R}}$, the CVC subsystem is activated, and GFMC operates following the $P_{\mathrm{c1}}^{\mathrm{CVC}}$-$\delta$ curve (the blue solid line). Otherwise, the CLC subsystem is activated, and its operating trajectory follows the $P_{\mathrm{c1}}^{\mathrm{CVC}}$-$\delta$ curve (the red solid line).

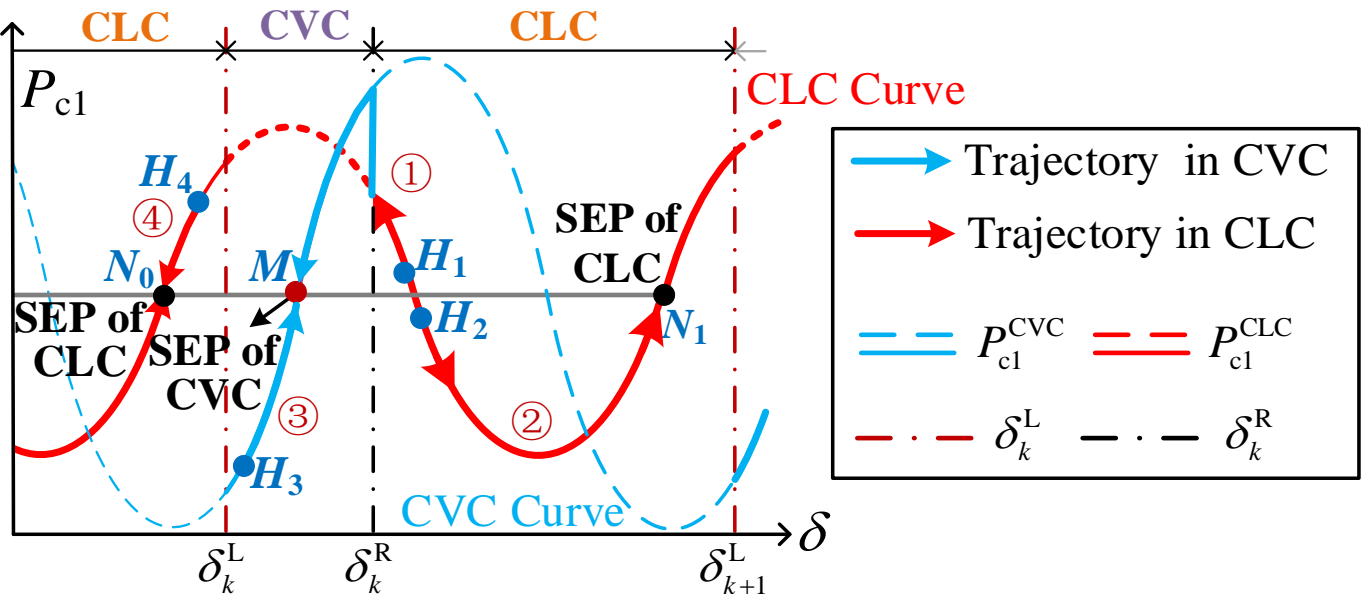

**Fig. 3.** GFMC angle-dependent Switching conditions on the $P_{\mathrm{c1}}$-$\delta$ plane

Eq.(8) shows that the PLL angle $\theta$ influences the GFMC's switching conditions through $\varphi_{\mathrm{c2}}$. If the PLL is converged, $\theta$ equals $\theta_0$. Here, $\theta_0$ is not an independent variable but rather a dependent variable that changes with $\delta$. Thus, $\delta_k^{\mathrm{L}}$ and $\delta_k^{\mathrm{R}}$ remain constant. However, if the PLL has not converged, $\delta_k^{\mathrm{L}}$ and $\delta_k^{\mathrm{R}}$ will fluctuate with $\theta$.

**2) GFLC angle-dependent Switching Conditions**

The GFLC angle-dependent switching condition for the fast variable $\theta$ is subsequently derived. According to Fig. 1(a), $U_{\mathrm{c1}}\angle\delta_0$ can be expressed as:

$$U_{\mathrm{c1}}\angle\delta_0 = U_{\mathrm{g}}\angle 0^\circ + \beta_1 I_{\mathrm{c1}}\angle\left(\varphi_{\mathrm{c1}} + 90^\circ\right) + \gamma I_{\mathrm{c2}}\angle\left(\varphi_{\mathrm{c2}} + 90^\circ\right), \tag{9}$$

where $\beta_1=1/|Y_{\mathrm{g}}|+1/|Y_{\mathrm{c1}}|$; $\gamma=1/|Y_{\mathrm{g}}|$. As per (9), $U_{\mathrm{c1}}\angle\delta_0$ is the sum of the grid voltage vector $U_{\mathrm{g}}\angle 0^\circ$ and the linearly transformed current vectors of both the GFMC and GFLC. This is valid for both the CVC and CLC subsystems, as shown in Fig. 4(a) and (b). However, they differ in that: in the CVC subsystem, the GFMC's voltage magnitude $U_{\mathrm{c1}}$ is maintained constant (meaning the length of the black vector in Fig. 4(a) is fixed); in contrast, for the CLC subsystem, the GFMC's maximum current amplitude $I_{\mathrm{c1}}^{\max}$ remains constant (meaning the length of the blue vector in Fig. 4(b) is fixed). Therefore, in the CVC subsystem, the terminal of $U_{\mathrm{c1}}\angle\delta$ traces an arc with a radius of $U_{\mathrm{c1}}$, forming the trajectory $\Gamma_{\mathrm{V}}$. By contrast, in the CLC subsystem, the endpoint of $\beta_1 I_{\mathrm{c1}}^{\max}\angle(\varphi_{\mathrm{c1}}+90^\circ)$ forms a circle centered at $(U_{\mathrm{g}},0)$ with a radius of $\beta_1 I_{\mathrm{c1}}^{\max}$, denoted as $\Gamma_{\mathrm{L1}}$.

The analysis below explains why PLL oscillation can cause system switching. In Fig. 4, the red vector, $\gamma I_{c2}\angle(\varphi_{c2}+90°)$, has a linear relationship with $I_{c2}\angle\varphi_{c2}$. Due to $\varphi_{c2}=\theta+\eta_2$, this red vector rotates along with the PLL oscillation. In the CVC subsystem, as the red vector rotates, the magnitude and angle of the blue vector adjust to maintain a constant magnitude for $U_{c1}\angle\delta$, thus preserving CVC mode. However, if the red vector rotates to a position where the endpoint of $U_{c1}\angle\delta$ cannot fall on $\Gamma_V$ irrespective of the blue vector's variation within $\Gamma_{L1}$, the system will then switch from CVC to CLC.

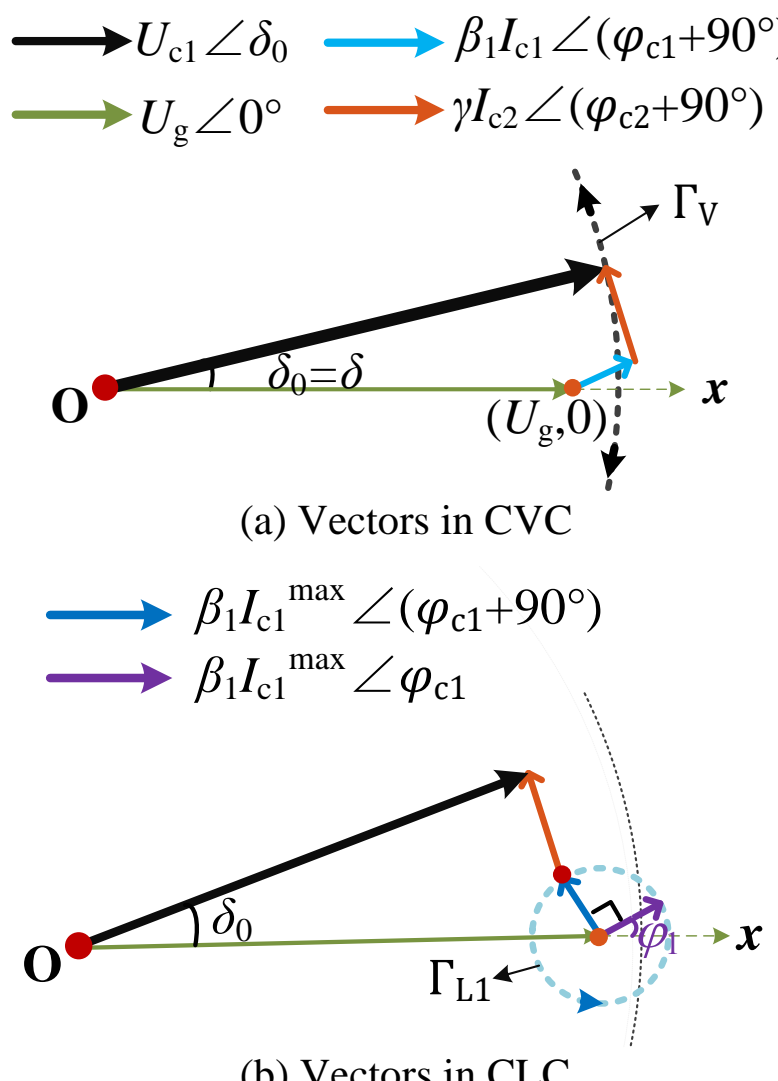

(a) Vectors in CVC

(b) Vectors in CLC

**Fig. 4.** Phasor diagrams in CVC and CLC subsystems.

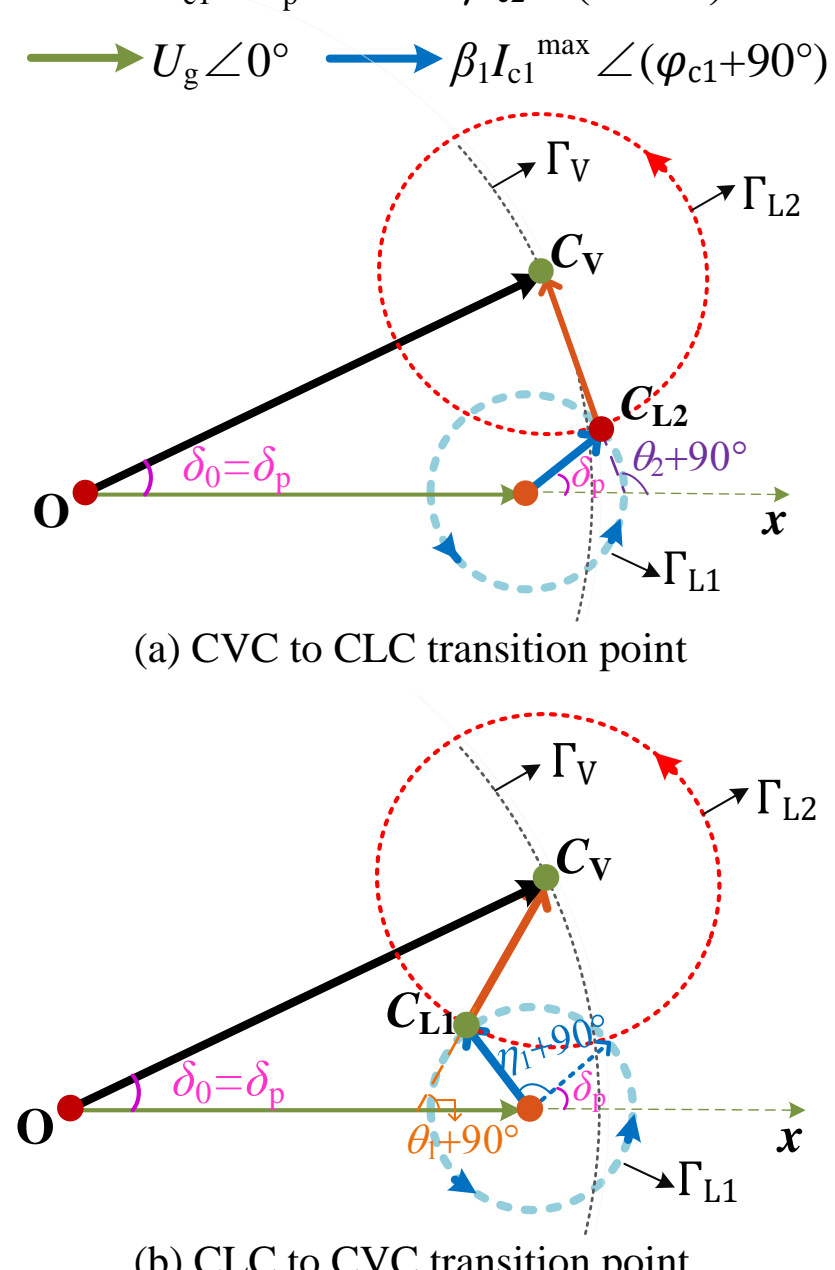

(a) CVC to CLC transition point

(b) CLC to CVC transition point

**Fig. 5.** Phasor diagram of GFLC angle-dependent Switching conditions.

Given the dynamic decoupling of slow variable $\delta$ and fast variable $\theta$, we fix $\delta=\delta_p$ to identify the critical $\theta$ values at the switching instant. After fault clearance, all the phasors change continuously until the current saturation angle undergoes an abrupt change (typically following current magnitude saturation). Thus, the phasor diagrams are identical at the pre- and post-switching instants ($t_-$ and $t_+$), provided this discontinuity has not yet occurred. Furthermore, since $\delta$ is continuous, the following relationships hold for GFMC angles:

**1) For transitions from CVC to CLC:**

- The pre-switching voltage angle aligns with $\delta_p$: $\delta(t_-)=\delta_p$.
- The post-switching current angle, after subtracting the current factor angle $\eta_1$, aligns with $\delta_p$: $\delta(t_+)=\delta_p$.
- Consequently, $\delta(t_-)=\delta(t_+)-\eta_1=\delta_p$.

**2) For transitions from CLC to CVC:**

- The pre-switching current angle, after subtracting the current factor angle $\eta_1$, aligns with $\delta_p$: $\delta(t_-)=\delta_p$.
- The post-switching voltage angle aligns with $\delta_p$: $\delta(t_+)=\delta_p$.
- Consequently, $\delta(t_-)-\eta_1=\delta(t_+)=\delta_p$.

Based on these conditions, the phasor relationships for CVC-to-CLC switching are shown in Fig. 5(a), and for CLC-to-CVC switching in Fig. 5(b). Let $\theta_1$ and $\theta_2$ denote the PLL angles at the CLC → CVC and CVC → CLC switching instants, respectively. From the geometric relationships, we can solve for:

$$\begin{cases} \theta_1 = 2\arccos\dfrac{\left(U_{c1}^{\ 2}-\beta_1 I_{c1}^{max}\right)^2+U_g^{\ 2}-\left(\gamma I_{c2}\right)^2}{2U_g\left(U_{c1}^{\ 2}-\beta_1 I_{c1}^{max}\right)}-\theta_2+\eta_1 \\ \theta_2 = \dfrac{\pi}{2}-\arccos\dfrac{\left(\gamma I_{c2}\right)^2+\left(U_g\right)^2-\left(U_{c1}^{\ 2}-\beta_1 I_{c1}^{max}\right)^2}{2\gamma I_{c2}U_g} \end{cases},\tag{10}$$

Thus, the mathematical model of the switching system is established. Eq.(6) serves as the system's governing dynamic equation. Eqs.(8) and (10) specify the GFMC angle-dependent switching condition and the GFLC angle-dependent switching condition of GFMC, respectively.

## III. Transient Stability Analysis of Switched System

This section analyzes the transient stability of this switched system, examining its correct SEPs, the transient stability of each subsystem, and the impact of switching processes.

### *A. Stable Equilibrium Points of the Switched System*

Define ($\theta_s^V$, 0) and ($\theta_s^L$, 0) as the SEPs for the CVC and CLC subsystems, respectively. Herein,

$$\theta_s^V=\arcsin\frac{\omega_g L_v i_{c2}^{d_2}}{U_{c1}}+\delta\ ,\tag{11}$$

$$\theta_s^L=\arcsin\frac{\omega_g L_g i_{c1}^{d_2}+\omega_g\left(L_g+L_{c2}\right)i_{c2}^{d_2}}{U_g}\ .\tag{12}$$

After the fault is removed, it is expected not only that the system will be stable, but also that it will stabilize at the SEP of the CVC subsystem rather than that of the CLC subsystem. Fig. 3 indicates that the system's ultimate stabilization location depends on the operating point's position at the instant of fault clearance. If operating points at fault clearance are $H_1$ and $H_3$, the system can return to the SEP of the CVC subsystem (point $M$) along trajectories ① and ③, respectively. However, if points are $H_2$ and $H_4$, the system will stabilize in the CLC subsystem along trajectories ② and ④, respectively. Therefore, if the system is in the CLC subsystem at the moment of fault clearance, there is a risk of it settling at the undesired SEP.

*B. Comparison of Two Subsystems' Stability Boundaries*

The transient stability boundaries of the switched system in the CVC and CLC subsystems are comparatively analyzed next.

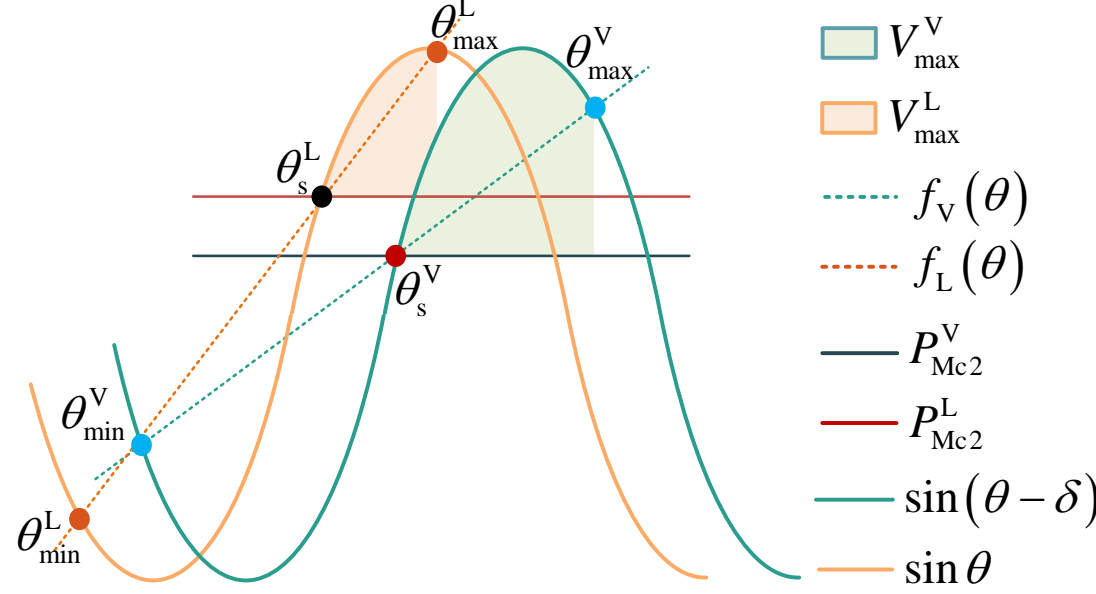

(a) Stability regions in CVC and CLC subsystem

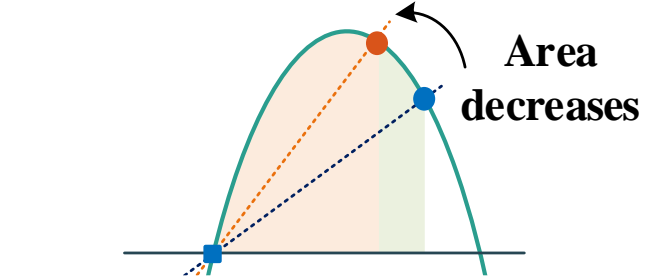

(b) Impact of increased slope on transient energy

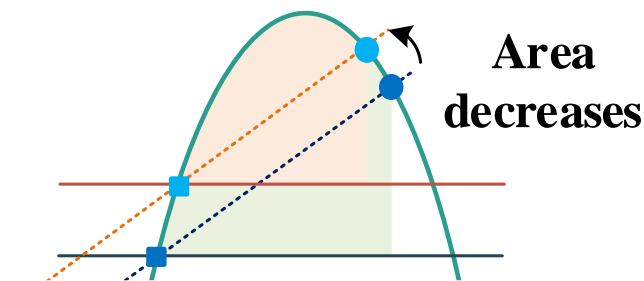

(c) Impact of increased intercept on transient energy

**Fig. 6.** Critical energy comparison between CVC and CLC subsystem.

The PLL's Lyapunov function can be chosen as[25]:

$$V(\theta,\varpi)=\frac{\varpi^{2}}{2}+\frac{1}{T_{c2}^{\{V,L\}}}\int_{\theta_{s}^{\{V,L\}}}^{\theta}\left(P_{Ec2}^{\{V,L\}}-P_{Mc2}^{\{V,L\}}\right)\mathrm{d}\theta\,, \tag{13}$$

where the superscript {V, L} denotes variables either in the CVC or CLC subsystems.

Ignoring the second-order damping in (6), the time derivation of (13) is d$V$/d$t$=-$D_{C2}^{\{V,L\}}\varpi^{2}$. The equilibrium point is stable if d$V$/d$t$ is non-positive definite. This is equivalent to the energy increment over one cycle (denoted as Δ$V$) being less than 0, which further implies that:

$$\Delta V=-\int_{0}^{t}\varpi^{2}D_{c2}^{\{V,L\}}\mathrm{d}t=-\oint\varpi D_{c2}^{\{V,L\}}\mathrm{d}\theta<0\Leftrightarrow\begin{cases}\int_{\theta_{s}}^{\theta}D_{c2}^{\{V,L\}}\mathrm{d}\theta>0,\theta>\theta_{s}\\ \int_{\theta_{s}}^{\theta}D_{c2}^{\{V,L\}}\mathrm{d}\theta<0,\theta<\theta_{s}\end{cases} \tag{14}$$

By substituting the expressions for $D_{c2}^{V}$ and $D_{c2}^{L}$ into (14), the condition for Δ$V$<0 is derived as:

$$\begin{aligned}&\text{In CVC subsystem}\begin{cases}\sin(\theta-\delta)>f_{V}(\theta),\ \text{if}\ \theta>\theta_{s}^{V}\\ \sin(\theta-\delta)<f_{V}(\theta),\ \text{if}\ \theta<\theta_{s}^{V}\end{cases}\\ &\text{In CLC subsystem}\begin{cases}\sin\theta>f_{L}(\theta),\ \text{if}\ \theta>\theta_{s}^{L}\\ \sin\theta<f_{L}(\theta),\ \text{if}\ \theta<\theta_{s}^{L}\end{cases}\end{aligned}\,, \tag{15}$$

where linear functions $f_V(\theta)$ and $f_L(\theta)$ are expressed as

$$\begin{cases}f_{V}(\theta)=\sin\left(\theta_{s}^{V}-\delta\right)+\dfrac{k_{2I}L_{v}}{k_{2p}U_{c1}}i_{c2}^{d_2}\left(\theta-\theta_{s}^{V}\right)\\ f_{L}(\theta)=\sin\theta_{s}^{L}+\left[\dfrac{k_{2I}\left(L_{g}+L_{c2}\right)}{k_{2p}U_{g}}i_{c2}^{d_2}+\dfrac{k_{2I}L_{g}}{k_{2p}U_{g}}i_{c1}^{d_2}+\dfrac{\omega_{g}L_{g}}{U_{g}}i_{c1}^{q_2}\right]\left(\theta-\theta_{s}^{L}\right)\end{cases} \tag{16}$$

Eq.(15) has clear geometric significance. A piecewise comparison between the sine function and the linear function can determine the size of the stability region. As illustrated in Fig. 6(a), the absolute stability region for the CVC subsystem is $\theta_{\min}^{V}<\theta<\theta_{\max}^{V}$, whereas for the CLC subsystem, it is $\theta_{\min}^{L}<\theta<\theta_{\max}^{L}$. In other words, the PLL is considered stable in CVC (CLC) if $\theta$ does not exceed $\theta_{\max}^{V}$ ($\theta_{\max}^{L}$) when $\varpi$=0. The critical energies for stability in the CVC and CLC subsystems are found by substituting $\theta_{\max}^{V}$ and $\theta_{\max}^{L}$ into (13), yielding $V_{\max}^{V}$ and $V_{\max}^{L}$, respectively. Numerically, $V_{\max}^{V}$ and $V_{\max}^{L}$ correspond to the green and red shaded areas in Fig. 6(a). A comparison of these values is then presented.

Fig. 6(b) and (c) demonstrate that increasing the slope and intercept of the linear function reduces the critical energy. In large-scale renewable energy sending end systems, the GFLC typically acts as the primary power source, with the GFMC serving as supporting equipment. Consequently, $i_{c2}^{d_2}$ is larger than $i_{c1}^{d_2}$ and $i_{c1}^{q_2}$. This means the slope of $f_L$ in (16) is primarily dictated by the term containing $i_{c2}^{d_2}$. Given that $L_g$+$L_{c2}$>$L_V$=$L_{c1}$+(1-$\alpha$)·$L_g$, the slope of $f_L$ is steeper than that of $f_V$. Furthermore, since $V_{\max}^{V}$ is independent of $\delta$ according to (11), (15), and (16), we can set $\delta$=0 without loss of generality. Under this condition, and because $P_{Mc2}^{L}$>$P_{Mc2}^{V}$, the intercept of $f_L$ exceeds that of $f_V$. Therefore, as illustrated in Fig. 6(a), the system stability in the CLC subsystem is significantly weaker than in the CVC subsystem.

*C. Impact of GFMC Switching on Transient Stability*

Given that the PLL's dynamics are much faster than that of PSL, GFMC angle-dependent switching has a limited impact on the PLL dynamic timescale. Its primary effect is instead reflected in the system's initial state at fault clearance, determining whether it is in the CVC or CLC subsystem. Therefore, the subsequent analysis will focus on the impact of GFLC angle-dependent switching, which incorporates the interplay between PLL dynamics and GFMC switching.

Analysis in Section II.C shows that in the PLL dynamic timescale, the CVC subsystem activates at $\theta$=$\theta_1$ and the CLC subsystem activates at $\theta$=$\theta_2$. Notably, $\theta_1$ and $\theta_2$ maintain a specific relationship with $\theta_{s}^{V}$, namely $\theta_1<\theta_{s}^{V}<\theta_2$. The detailed proof is in Appendix B. Next, we analyze how the switching dynamic affects transient energy changes. Fig. 7 shows the $P_{Ec2}^{V}$-$\theta$ curve (red) and $P_{Ec2}^{L}$-$\theta$ curve (blue). Point *B'* in the figure has an angle of $\theta_{s}^{V}$, while points *E* and *F* have angles $\theta_1$ and $\theta_2$, respectively. A switch from the CLC to the CVC subsystem is represented by the transition from *E* to *E'*. Similarly, switching from the CVC to the CLC subsystem is depicted as a transition from *F* to *F'*.

The system's transient energy in (13) consists of two components: equivalent kinetic energy and equivalent potential energy, which correspond to the first and second terms on the right-hand side of the equation, respectively. As depicted in the PLL's control scheme in Fig. 1(c), $\varpi$ is the output signal of the PI controller. Since the proportional gain ($k_{2p}$) is typically much smaller than the integral gain ($k_{2I}$), $\varpi$ is predominantly determined by the integral action. Due to the inherent lag of the integrator, $\varpi$ remains approximately constant at the instant of system switching. Consequently, the kinetic energy is virtually unchanged immediately before and after switching. Thus, any change in the system's total energy across the switching instant is primarily due to changes in its potential energy.

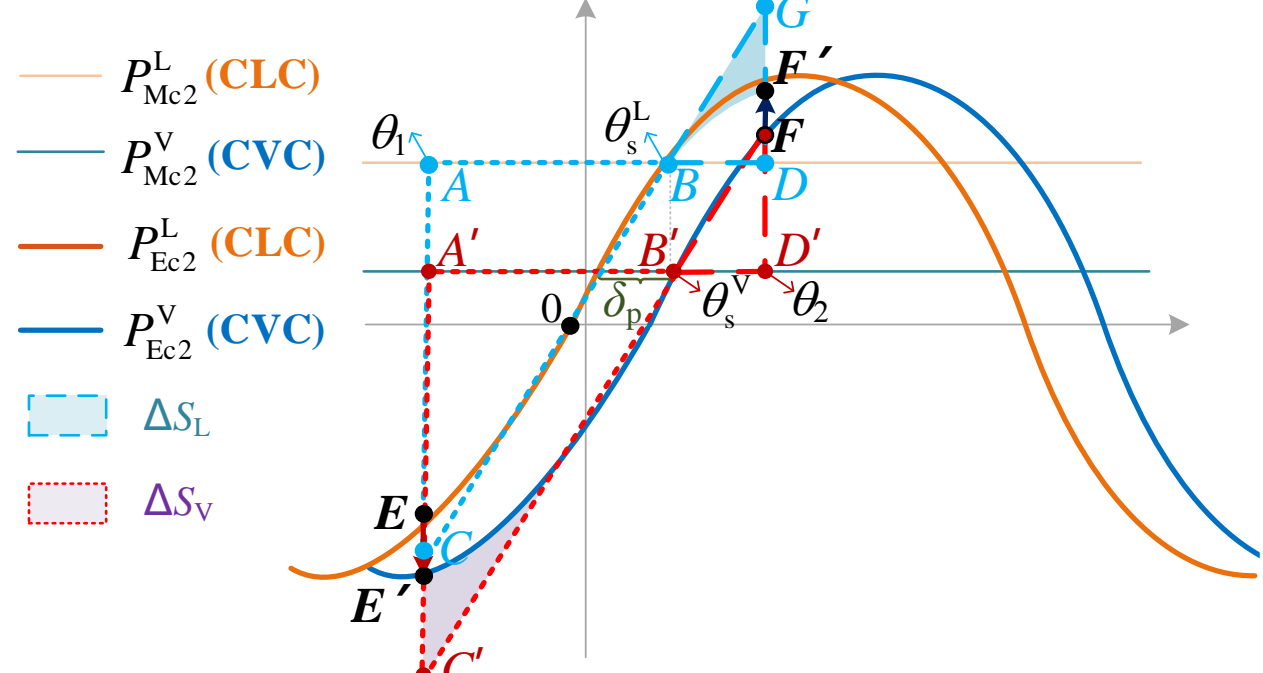

**Fig. 7.** Switching dynamics on the $P_{\mathrm{Ec2}}^{\{\mathrm{V,L}\}}$-$\theta$ plane

Let $\Delta V_{\mathrm{o}}$ denote the overall periodic energy increment for the full switching sequence from CLC to CVC and then back to CLC. The potential energy discontinuities at points $E$ and $F$ are represented by $\Delta V_{\mathrm{p}}^{E}$ and $\Delta V_{\mathrm{p}}^{F}$, respectively. The periodic energy change attributable to damping is denoted by $\Delta V_{\mathrm{D}}$. Thus,

$$\begin{aligned}\Delta V_{\mathrm{o}}=&\underbrace{\frac{1}{T_{\mathrm{c2}}^{\mathrm{V}}}\int_{\theta_{\mathrm{s}}^{\mathrm{V}}}^{\theta_1}\left(P_{\mathrm{Ec2}}^{\mathrm{V}}-P_{\mathrm{Mc2}}^{\mathrm{V}}\right)\mathrm{d}\theta-\frac{1}{T_{\mathrm{c2}}^{\mathrm{L}}}\int_{\theta_{\mathrm{s}}^{\mathrm{L}}}^{\theta_1}\left(P_{\mathrm{Ec2}}^{\mathrm{L}}-P_{\mathrm{Mc2}}^{\mathrm{L}}\right)\mathrm{d}\theta}_{\Delta V_{\mathrm{p}}^{E}}\\&+\underbrace{\frac{1}{T_{\mathrm{c2}}^{\mathrm{L}}}\int_{\theta_{\mathrm{s}}^{\mathrm{L}}}^{\theta_2}\left(P_{\mathrm{Ec2}}^{\mathrm{L}}-P_{\mathrm{Mc2}}^{\mathrm{L}}\right)\mathrm{d}\theta-\frac{1}{T_{\mathrm{c2}}^{\mathrm{V}}}\int_{\theta_{\mathrm{s}}^{\mathrm{V}}}^{\theta_2}\left(P_{\mathrm{Ec2}}^{\mathrm{V}}-P_{\mathrm{Mc2}}^{\mathrm{V}}\right)\mathrm{d}\theta}_{\Delta V_{\mathrm{p}}^{F}}+\Delta V_{\mathrm{D}}\end{aligned}\quad.(17)$$

Appendix C demonstrates that the relationship $\theta_{\mathrm{s}}^{\mathrm{V}}\leqslant\theta_{\mathrm{s}}^{\mathrm{L}}$ holds at the system switching instant. Fig. 7 illustrates the specific case where $\theta_{\mathrm{s}}^{\mathrm{V}}=\theta_{\mathrm{s}}^{\mathrm{L}}$. From (20), the potential energies at points $E$, $E'$, $F$, and $F'$ are equal to the areas of regions $ABE$, $A'B'E'$, $BDF'$, and $B'D'F$, which are denoted as $S_{ABE}$, $S_{A'B'E'}$, $S_{BDF'}$, and $S_{B'D'F}$, respectively. These areas are enclosed by boundary segments formed from sections of curves $P_{\mathrm{Ec2}}^{\{\mathrm{V,L}\}}$, $P_{\mathrm{Mc2}}^{\{\mathrm{V,L}\}}$, and $\theta=\theta_{\mathrm{s}}^{\{\mathrm{V,L}\}}$, $\theta=\theta_1$, and $\theta=\theta_2$. Key points (e.g., $A$, $B$, $E$ for region $ABE$) serve to identify the specific enclosed region by marking the vertices or extents of these boundary segments. As illustrated in Fig. 7, region $B'D'F$, which is enclosed by $P_{\mathrm{Ec2}}^{\mathrm{V}}$, $P_{\mathrm{Mc2}}^{\mathrm{V}}$, and $\theta=\theta_2$, can be approximated by triangle $B'D'F$, such that $S_{B'D'F}\approx S_{\Delta B'D'F}$. Triangle $BDG$ is constructed to be congruent to triangle $B'D'F$, thus $S_{\Delta BDG}=S_{\Delta B'D'F}$. Since the $P_{\mathrm{Ec2}}^{\mathrm{L}}$-$\theta$ curve exhibits greater concavity than the $P_{\mathrm{Ec2}}^{\mathrm{V}}$-$\theta$ curve at the switching point $\theta=\theta_2$, it is evident that $S_{\Delta BDG}>S_{BDF'}$, with their difference denoted as $\Delta S_{\mathrm{L}}$. Therefore, the potential energy at point $F$ is less than that at point $F'$ (i.e., $\Delta V_{\mathrm{p}}^{F}<0$), indicating system energy decay during the switch from CVC to CLC. Similarly, it can be known that switching from CLC to CVC also results in energy decay. To further clarify, if $\theta_{\mathrm{s}}^{\mathrm{V}}<\theta_{\mathrm{s}}^{\mathrm{L}}$, this area comparison method readily shows that this energy reduction becomes even more pronounced. Thus, the GFLC angle-dependent switching process leads to energy decay within the switched system.

In summary, the CLC subsystem exhibits significantly weaker stability compared to the CVC subsystem. The GFLC angle-dependent switching contributes to transient energy dissipation. As a result, the stability boundary of the switched system is determined by the CLC subsystem. The critical energy can be calculated by substituting $\theta_{\mathrm{max}}^{\mathrm{L}}$ into (13).

## IV. Stability Enhancement Control

This section designs a method for enhancing the stability of PLLs and stabilizing the PSL to CVC rather than CLC, solely based on feedback from each converter's own state variables.

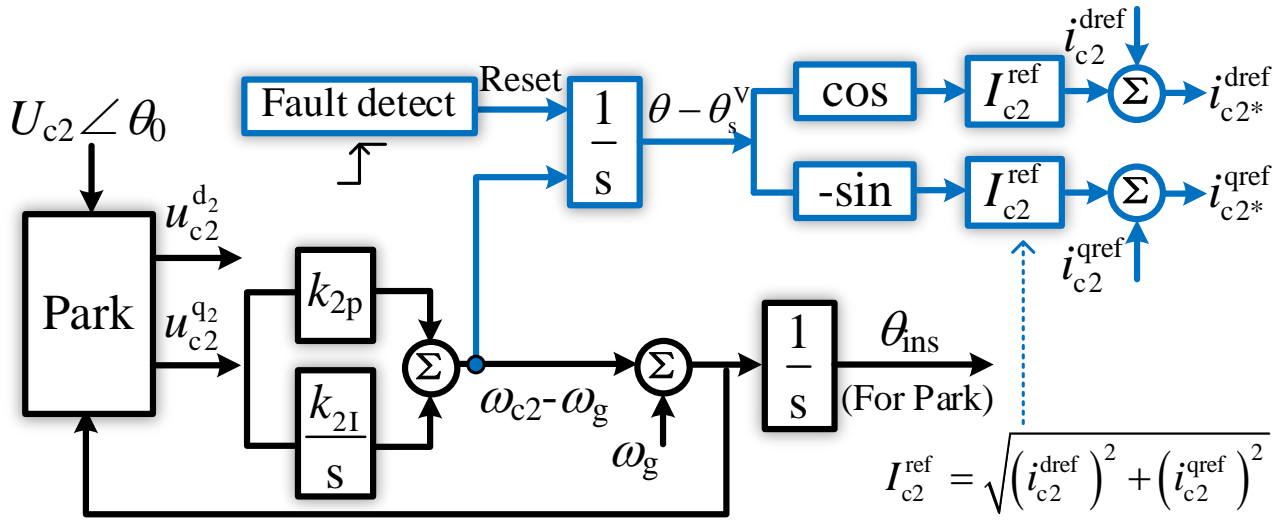

(a) GFLC control structure with VFDC

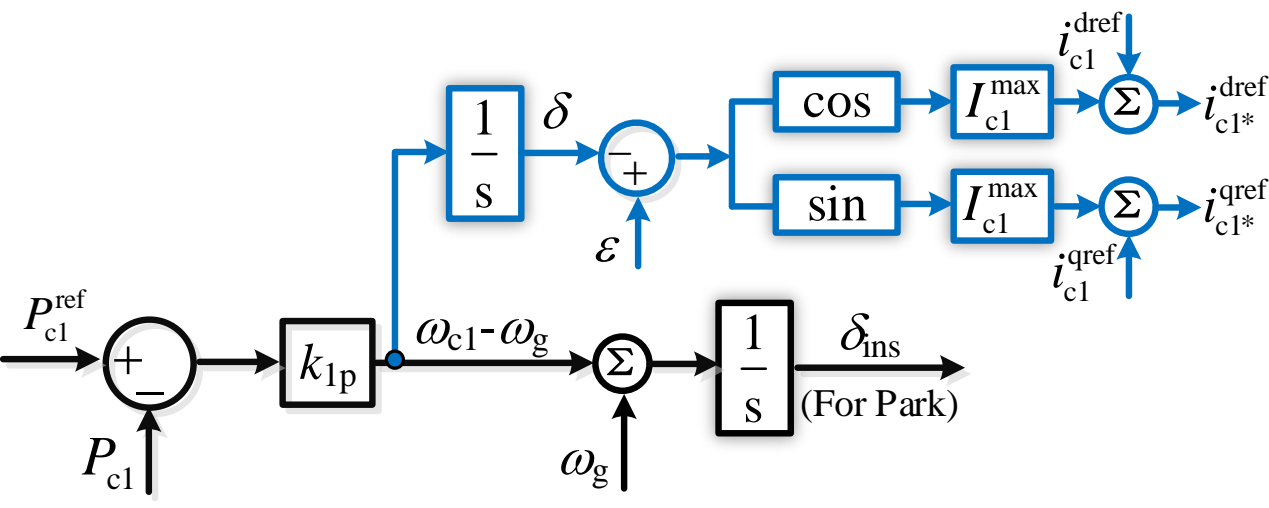

(b) GFMC control structure with VFDC

**Fig. 8.** Control block diagram of VFDC.

### *A. Virtual d-Axis Fixed Control Strategy*

Under traditional grid-following control of GFLC or the CLC control of GFMC, the converter's d-axis is typically aligned with $\theta$ or $\delta$. This alignment keeps the d-axis component of the GFLC or GFMC current constant despite variations in $\theta$ or $\delta$. Consequently, as $\theta$ increases, this constant current component can provide sustained "accelerating torque", threatening PLLs' transient stability. Moreover, this control approach carries the risk of the GFMC remaining in CLC mode after fault clearance, as analyzed above. Furthermore, it may lead to interactive current coupling between converters, thereby increasing the difficulty of control.

To address this, we introduce the concept of "Virtual Fixed d-axis Control (VFDC)". This strategy involves two key steps: first, each converter's virtual d-axis is aligned with the actual current vector. Then, through each converter's own state feedback, the current vector angle relative to the synchronous rotating axis is maintained constant.

The specific implementation is simple. For each grid-following converter, let:

$$\eta_2=\theta_{\mathrm{s}}^{\mathrm{V}}-\theta\,. \tag{18}$$

For centralized grid-connected droop-controlled GFMCs, let:

$$\eta_1=\varepsilon-\delta\,. \tag{19}$$

where $\varepsilon$ is a constant. The control block diagram for this approach is presented in Fig. 8.

As a result of implementing (18) and (19), the virtual d-axis angles (current vector angles) for the GFMC and GFLC, $\varphi_{c1*}$ and $\varphi_{c2*}$, are now, respectively:

$$\varphi_{c1*} = \varepsilon,\ \varphi_{c2*} = \theta_{s}^{V}\,. \tag{20}$$

As depicted in Fig. 9, with VFDC applied, the converter's current angle no longer tracks $\theta$ or $\delta$. Instead, it maintains a relatively constant angle with respect to the synchronous rotating axis (x-axis). Specifically, this constant angle is $\theta_{s}^{V}$ for GFLC and $\varepsilon$ for GFMC. This achieves dynamic decoupling among multiple GFLCs, and also between GFLCs and GFMC.

The following discussion will elaborate on how this decoupling approach enhances system stability and facilitates adaptive stabilization to the correct equilibrium point.

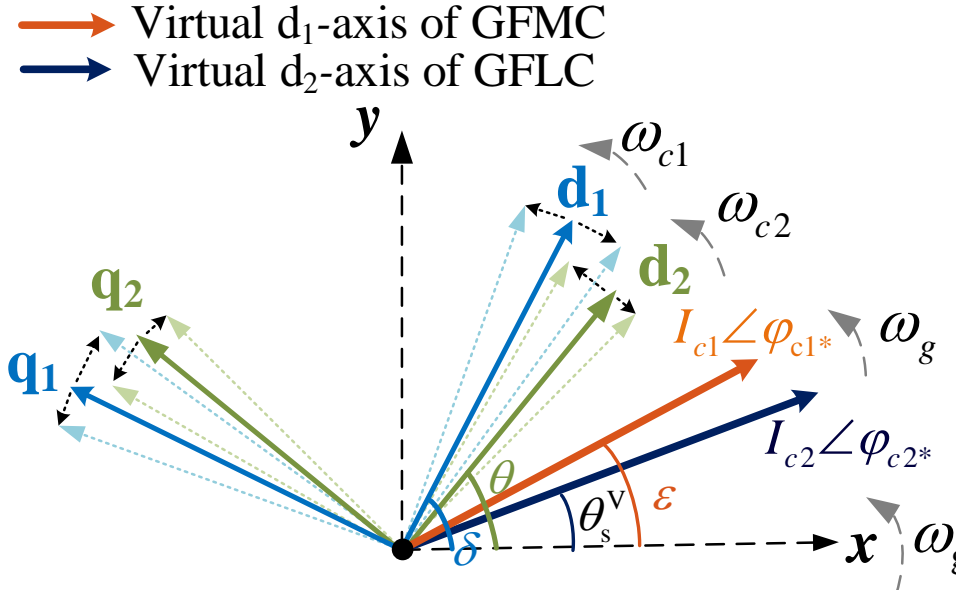

**Fig. 9.** System vectors after applying the proposed VFDC

*B. Demonstration of Stability Enhancement*

Subsequent analysis will show that for co-located GFLCs and droop-controlled GFMCs systems, VFDC achieves superior transient stability over conventional current control of GFLCs and GFMCs.

**1) Demonstration of enhanced PLL stability**

In engineering practice, renewable energy stations typically incorporate multiple GFLCs operating in parallel, often with diverse control parameters and power outputs due to varying locations. In contrast, GFMCs used for support can be centrally fed in and uniformly controlled. Consequently, it is need to analyze scenarios with multiple GFLCs. Non-coherent GFLCs operation can induce additional negative damping via coupling [26], potentially compromising transient stability relative to coherent scenarios. Consequently, if the system's stability with VFDC proves superior to that of multi-GFLC coherent operation, the stability region achieved with VFDC will inevitably exceed that under non-coherent conditions.

Let the subscript "$i$" denote the $i$-th GFLC (with $n$ being the total number). $D_{c2}^{V}$ and $D_{c2}^{L}$ are updated to:

$$\begin{cases} D_{c2i}^{V} = \dfrac{k_{2pi}}{k_{2Ii}}\left[\alpha U_{c1}\cos\left(\theta_i-\delta\right)+\left(1-\alpha\right)U_{g}\sin\theta_i\right]-L_{vi}i_{c2}^{d_2}-\left(1-\alpha\right)L_{g}i_{c2ij}^{d_2} \\ D_{c2i}^{L} = \dfrac{k_{2pi}}{k_{2Ii}}\left(U_{g}\cos\theta_i+\omega_{g}L_{g}i_{c1}^{q_2}\right)-L_{g}i_{c1}^{d_2}-\left(L_{g}+L_{c2i}\right)i_{c2i}^{d_2}-L_{g}i_{c2ij}^{d_2} \end{cases} \tag{21}$$

where $i_{c2ij}^{d_2}$ is the current coupling term between the $i$-th GFLC and the other GFLCs, given by

$$i_{c2ij}^{d_2} = \sum_{j=1,\, j\neq i}^{n} I_{c2j}\cos\left(\theta_i-\varphi_{c2j}\right) \tag{22}$$

and $L_{vi}$=$L_{c2i}$+(1-$\alpha$)$L_{g}$. Substituting (21) into (14) yields the updated linear functions $f_{V}(\theta_i)$ and $f_{L}(\theta_i)$ as:

$$\begin{cases} f_{V}\left(\theta_i\right)=\sin\left(\theta_{si}^{V}-\delta\right)+\dfrac{k_{2Ii}}{k_{2pi}U_{c1}}\left[L_{v}i_{c2i}^{d_2}+\left(1-\alpha\right)L_{g}i_{c2ij}^{d_2}\right]\left(\theta_i-\theta_{si}^{V}\right) \\ f_{L}\left(\theta_i\right)=\sin\theta_{si}^{L}+\left[\dfrac{k_{2Ii}\left(L_{g}+L_{c2i}\right)}{k_{2pi}U_{g}}i_{c2i}^{d_2}+\dfrac{k_{2Ii}L_{g}}{k_{2pi}U_{g}}i_{c2ij}^{d_2}+\dfrac{k_{2Ii}L_{g}}{k_{2p}U_{g}}i_{c1i}^{d_2}-\dfrac{\omega_{g}L_{g}}{U_{g}}i_{c1i}^{q_2}\right]\left(\theta_i-\theta_{si}^{L}\right) \end{cases} \tag{23}$$

Upon implementing VFDC, eq. (23) changes to:

$$\begin{cases} f_{V*}\left(\theta_i\right)=\sin\left(\theta_{si}^{V}-\delta\right)+\dfrac{k_{2Ii}}{k_{2pi}U_{c1}}\left[L_{v}I_{c2i}\cos\left(\theta_{si}^{V}-\theta_i\right)+\left(1-\alpha\right)L_{g}\sum\limits_{j=1,\, j\neq i}^{n}I_{c2j}\cos\left(\theta_{sj}^{V}-\theta_i\right)\right]\left(\theta_i-\theta_{si}^{V}\right) \\ f_{L*}\left(\theta_i\right)=\sin\theta_{si}^{L}+\left[\dfrac{k_{2Ii}\left(L_{g}+L_{c2i}\right)}{k_{2pi}U_{g}}I_{c2i}\cos\left(\theta_{si}^{V}-\theta_i\right)+\dfrac{k_{2Ii}L_{g}}{k_{2pi}U_{g}}\sum\limits_{j=1,\, j\neq i}^{n}I_{c2j}\cos\left(\theta_{sj}^{V}-\theta_i\right)+\dfrac{k_{2Ii}L_{g}}{k_{2pi}U_{g}}i_{c1i}^{d_2}-\dfrac{\omega_{g}L_{g}}{U_{g}}i_{c1i}^{q_2}\right]\left(\theta_i-\theta_{si}^{L}\right) \end{cases} \tag{24}$$

As per the analysis in Section III.B, a decrease in the slope of linear functions signifies an expansion of the stability boundary. After the fault is cleared, GFLCs revert to the normal control state. Thus, $I_{c2i}\cos(\theta_{s}^{V}-\theta)\leq I_{c2i}\approx i_{c2}^{d_2}$. Consequently, a comparison of (23) (without VFDC) and (24) (with VFDC) indicates that the slopes of the linear functions $f_{V}$ and $f_{L}$ without VFDC are greater than those of $f_{V*}$ and $f_{L*}$ with VFDC, as depicted in Fig. 10(a). This implies that VFDC expands the PLL stability boundary in both CVC and CLC subsystems. Furthermore, the slopes of $f_{V*}$ and $f_{L*}$ decrease with increasing $\theta$, leading to a dynamic expansion of the stability boundary.

**2) Demonstration of global stability of GFMC in CVC**

Sustained operation in CLC is unacceptable for GFMCs. Therefore, it is imperative that the GFMC ultimately returns to CVC, demonstrating robust adaptability to various preceding disturbances. The VFDC addresses this challenge by inducing a unidirectional drift in the GFMC's frequency and angle when in CLC mode. This drives the operating point out of the CLC region, allowing the converter to switch back to CVC and stabilize. The demonstration is as follows:

Upon applying VFDC, the expression for $P_{c1}^{CLC}$ (GFMC's active power in CLC) is updated to $P_{c1*}^{CLC}$:

$$P_{c1*}^{CLC} = k_{A}\sin\left(\varepsilon+\beta\right)$$

$$\text{with}\begin{cases} k_{A}=\sqrt{\left(U_{g}I_{c1}^{max}\right)^{2}+\left(\dfrac{I_{c1}^{max}I_{c2}}{|Y_{g}|}\right)^{2}-\dfrac{2U_{g}\left(I_{c1}^{max}\right)^{2}I_{c2}}{|Y_{g}|}\sin\theta_{s}^{V}} \\ \beta=\arctan\dfrac{|Y_{g}|U_{g}-I_{c2}\sin\theta_{s}^{V}}{I_{c2}\cos\theta_{s}^{V}} \end{cases}. \tag{25}$$

In (25), $k_{A}$ and $\beta$ can be considered constants during the transient period. As a result, $P_{c1*}^{CLC}$ forms a straight line parallel to the $\delta$-axis, as depicted in Fig. 10(b). If $P_{c1*}^{CLC}\neq P_{c1}^{CLC}$, a persistent (and intentionally introduced) imbalance $P_{c1*}^{CLC}$-$P_{c1}^{CLC}$ will always exist. According to (1), this imbalance will cause $\delta$ to continuously increase if $P_{c1*}^{CLC}>P_{c1}^{CLC}$ (and vice-versa). This continuous shift of $\delta$ in one direction persists until it reaches the switching boundary $\delta$=$\delta_{k}^{L}$ (or $\delta$=$\delta_{k}^{R}$). At this juncture, GFMC transitions from CLC to CVC mode. Graphically, as shown in Fig. 10(b), the operating point jumps from the blue dashed line to the purple curve. Subsequently, it moves along this curve until it converges to the SEP within the CVC region. Notably, VFDC's controlled guidance mechanism leads the GFMC to a viable CVC SEP, which may reside in a subsequent or preceding cycle. This ultimately ensures the global stability of the switched system in CVC.

To ensure safety, it is preferable for the active power to remain below its reference value. Thus, the subsequent step involves adjusting $\varepsilon$ such that $P_{c1*}^{CLC}<P_{c1}^{CLC}$. Given network constraints, $\beta$ is confined to the range [0, π/2]. As a result, $P_{c1*}^{CLC}$ exhibits a monotonic increase with $\varepsilon$ over the range $\varepsilon\in[-\pi/2,0]$. Therefore, choosing $\varepsilon=-\pi/2$ ensures that $P_{c1*}^{CLC}$ is indeed less than $P_{c1}^{CLC}$, and concurrently offers the most significant benefit to GFLCs' transient stability.

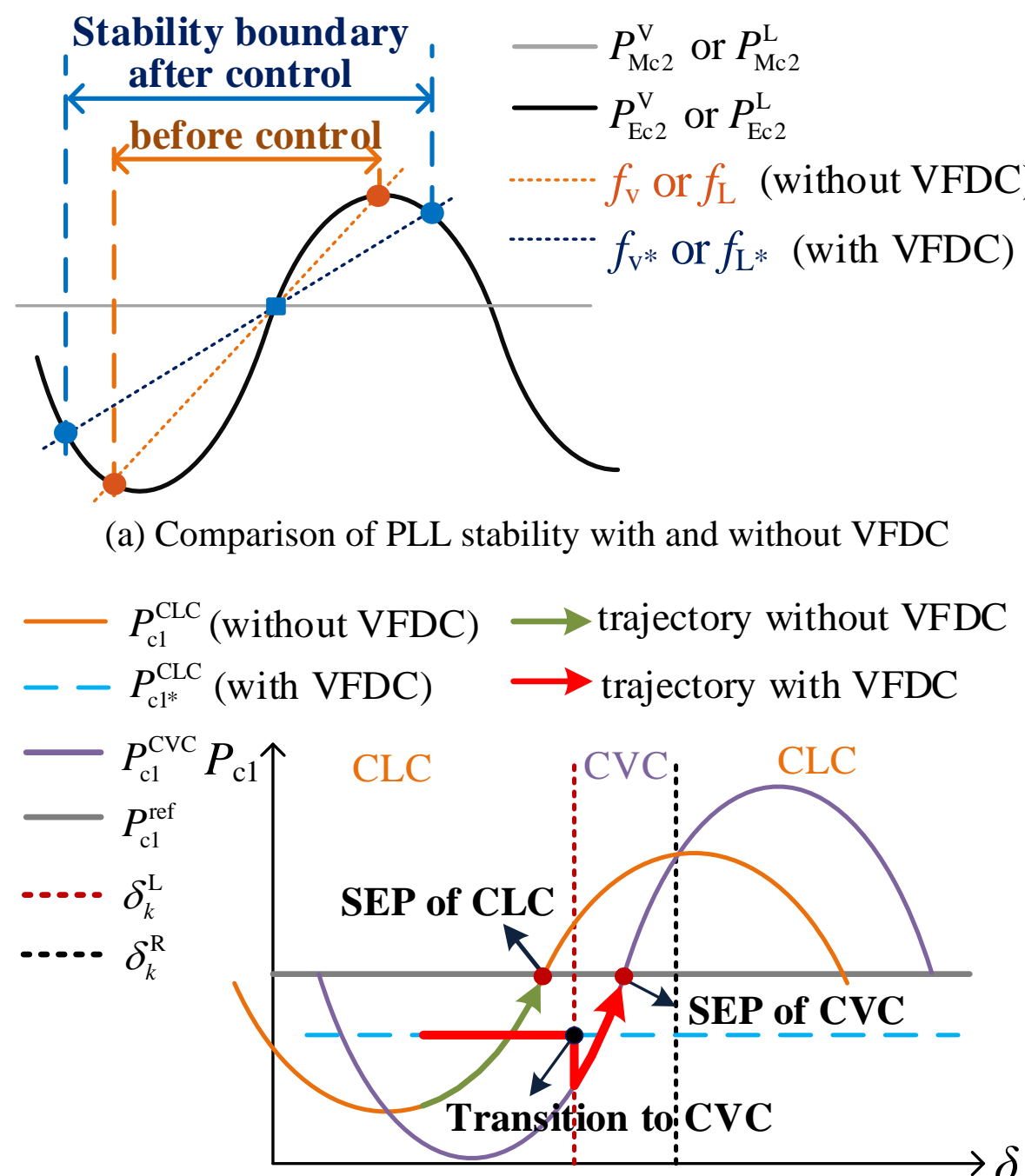

(a) Comparison of PLL stability with and without VFDC

(b) Comparison of GFMC trajectories with and without VFDC

**Fig. 10.** Illustration of the effects of VFDC.

Furthermore, for GFLCs, it is recommended that VFDC control be activated upon entering low-voltage ride-through (LVRT) control and sustained for several seconds. For GFMCs, VFDC control should be engaged if in CLC mode.

In summary, the proposed VFDC control strategy improves the GFLC's transient synchronization stability and drives the PSL-GFMC to adaptively return to the SEP of the CVC.

## V. Verification

### A. Test System 1: Electromagnetic Transient Simulation

A simplified equivalent model of the GFLC and PSL-GFMC co-located system is developed using the PSCAD simulation platform, as depicted in Fig. 1(a). The fault scenario involves a grid voltage sag occurring at $t$=1s, where the voltage drops to 0.3-0.5 p.u. and persists for $t_c$ seconds. TABLE III in Appendix C details the simulation parameters, and TABLE I specifies the test conditions for Cases 1 to 7.

#### 1) Verification of the transient stability analysis

Cases 1-3 are designed to verify the impact on switched system stability arising from two scenarios: sustained operation within either the CVC or CLC subsystems (representing GFMC angle-dependent switching), and rapid, alternating CVC/CLC activation driven by PLL oscillations (representing GFLC angle-dependent switching). The test conditions for Cases 1, 2, and 3 are largely consistent, with the primary distinction being the progressive increase in the GFMC's current limit ($I_{c1}^{max}$).

TABLE I
CONDITIONS AND PARAMETERS FOR TEST SYSTEM 1

| Case | $t_c$ (s) | $P_{c1}^{ref}$,$P_{c2}^{ref}$ (MW) | $K_q$ | $K_{2p}$ | $I_{c1}^{max}$ (p.u.) | $\eta_1$(rad) | $\rho$ |
|---|---|---|---|---|---|---|---|
| 1 | 0.05 | 35,110 | 3 | 0.07 | 1.1 | 0 | 0.3 |
| 2 | 0.05 | 35,110 | 3 | 0.07 | 1.25 | 0 | 0.3 |
| 3 | 0.05 | 35,110 | 3 | 0.07 | 1.7 | 0 | 0.3 |
| 4 | 0.1 | 35,70 | 3 | 0.15 | 1.2 | 0 | 0.3 |
| 5 | 0.1 | 35,70 | 0.5 | 0.15 | 1.2 | -π/2 | 0.3 |
| 6 | 0.15 | 20,100 | 3 | 0.15 | 1.1 | 0 | 0.5 |
| 7 | 0.15 | 20,100 | 0.5 | 0.15 | 1.1 | -π/2 | 0.5 |
| 8 | 0.2 | 40, 60 | 0 | 0.075 | 1.3 | 0 | 0.1 |

**Notes**: a) $t_c$: fault duration; b) $P_{c1}^{ref}$, $P_{c2}^{ref}$: steady state power of GFMC and GFLC; c) LVRT reactive support coefficient of GFLC; d) $k_{1p}$: proportional coefficient of PLL; e) $I_{c1}^{max}$: GFMC current limit threshold; f) $\eta_1$: saturation current angle of GFMC; g) $\rho$: grid voltage drop coefficient during faults.

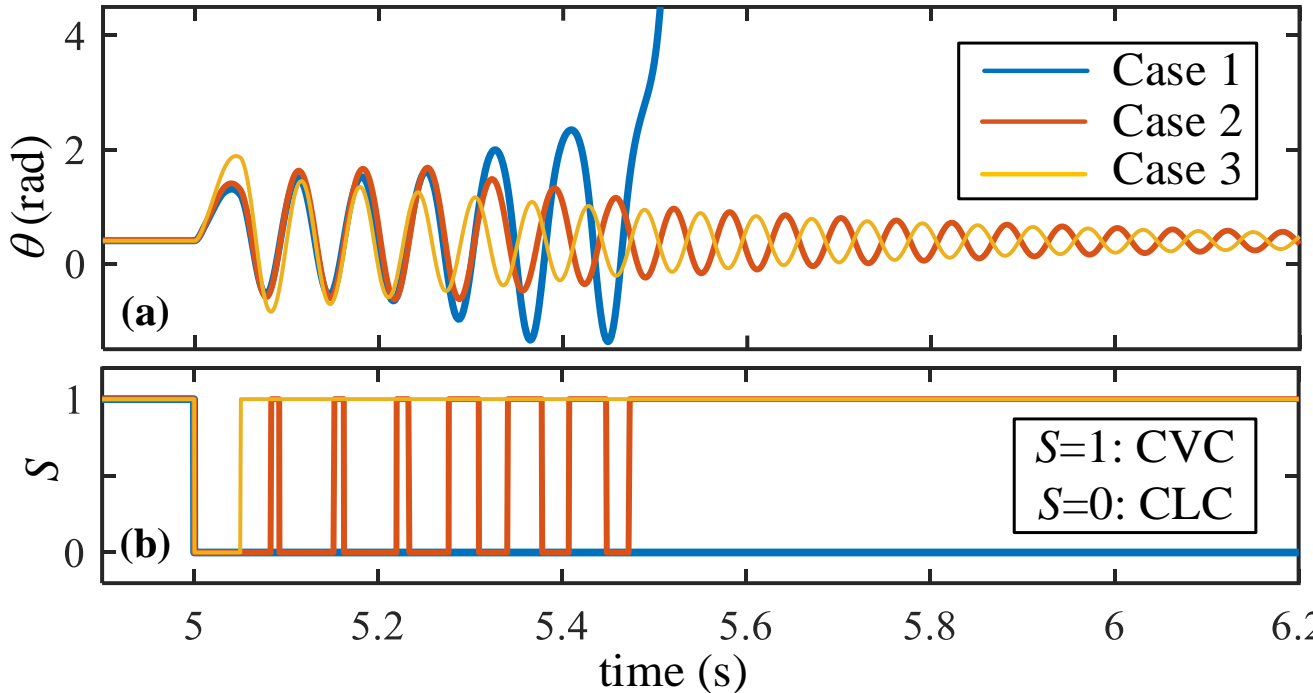

**Fig. 11.** Simulation Results of Cases 1, 2, and 3. (a) PLL angle versus time curves; (b) CVC and CLC state indicator.

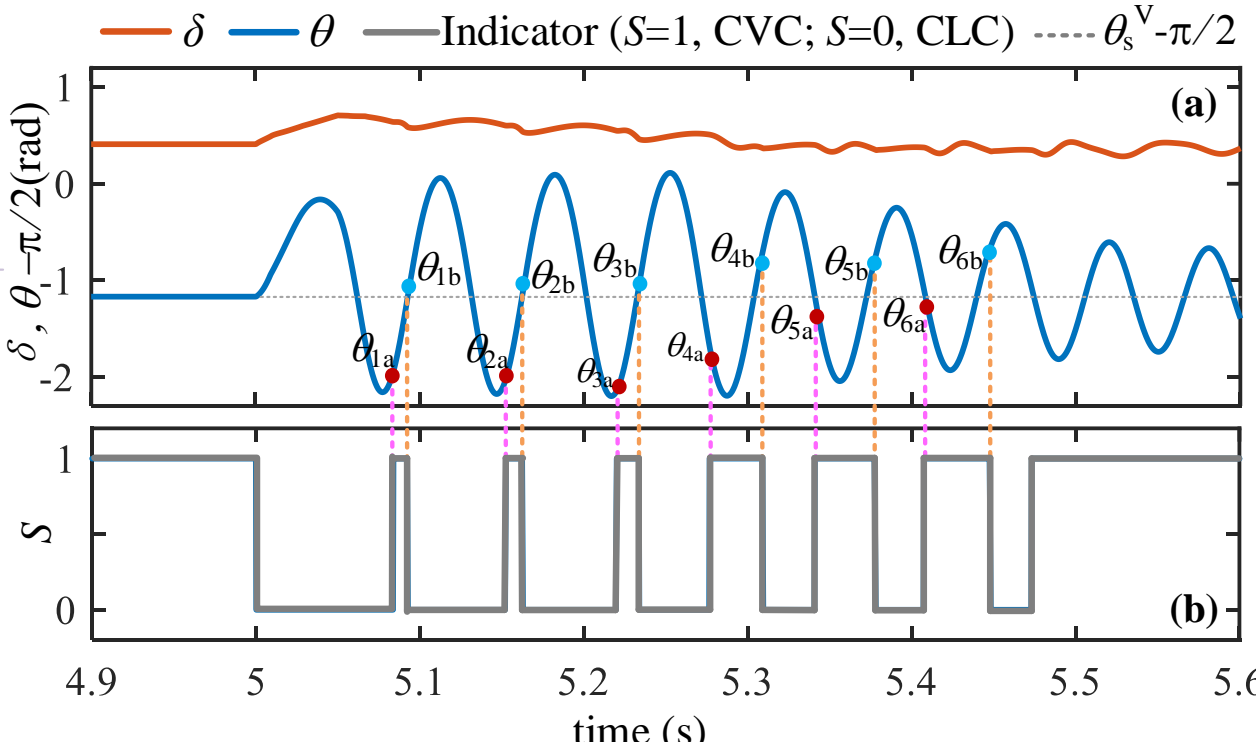

**Fig. 12.** Simulation Results of Case 2. (a) PSL and PLL angle versus time curves; (b) CVC and CLC state indicator.

The corresponding results are presented in Fig. 11 and Fig. 12. In Fig. 11 (b) and Fig. 12 (b), "$S$" serves as a switching indicator: $S$=1 denotes CVC subsystem activation, while $S$=0 indicates CLC subsystem activation. As depicted in Fig. 11 (b), the simulation results for Cases 1, 2, and 3 illustrate three distinct post-fault clearance scenarios: CLC activation (Case 1), alternating CLC and CVC activation (Case 2), and CVC activation (Case 3).

According to the transient energy calculations in Table II, the PLL's transient energy at the moment of fault clearance progressively increases from Case 1 to Case 3. This trend is attributed to the larger GFMC current amplitude within the fault period. As per (11), this large current increases $P_{Mc2}^{L}$, consequently leading to more significant acceleration during the fault. Despite this energy increase, comparing the PLL angle oscillation amplitudes in Fig. 11(a) for Cases 1 and 3 clearly shows the CVC subsystem's markedly superior stability over the CLC subsystem. This indicates a significantly larger stability domain boundary for the CVC subsystem, which validates the theoretical analysis presented in Section III.A.

Similarly, comparing the simulation results of Case 1 and Case 2 demonstrates that fast switching (Case 2) is more conducive to PLL stability than continuous CLC subsystem activation (Case 1). Fig. 12 provides a detailed depiction of the dynamic switching process in Case 2. As shown, during each PLL oscillation cycle, the variation in the PSL's angle remains relatively constant, allowing it to be regarded as a slow variable. In Fig. 12(a), $\theta_{ka}(+\pi/2)$ and $\theta_{kb}(+\pi/2)$ ($k$=1,…,7) denote the PLL angles at the $k$-th CVC subsystem activation and deactivation, respectively. The condition $\theta_{ka}<\theta_s^V<\theta_{kb}$ is observed, with the CVC subsystem activating at $\theta_{ka}$ (the left switching point) and the CLC subsystem activating at $\theta_{kb}$ (the right switching point). The energy calculation results in Table II further reveal that each CLC→CVC→CLC fast-switching cycle dissipates the system's transient energy. Therefore, the stability boundary of the switched system is ultimately determined by its stability within the CLC subsystem. These findings validate the theoretical analysis in Section III.B.

TABLE II
RESULTS OF TRANSIENT ENERGY CALCULATION

| Transient Energy | Value | Result |
|---|---|---|
| At fault clearance | Case 7: 0.65; Case 8: 0.72; Case 9: 1.33 | $V_{Case7}<V_{Case8}<V_{Case9}$ |
| In CLC at $\theta_{1a}$,$\theta_{1b}$ | $V(\theta_{1a})$: 0.61; $V(\theta_{1b})$: 0.14 | $V(\theta_{1b})<V(\theta_{1a})$ |
| In CLC at $\theta_{2a}$, $\theta_{2b}$ | $V(\theta_{2a})$: 0.77; $V(\theta_{2b})$: 0.16 | $V(\theta_{2b})<V(\theta_{2a})$ |
| In CLC at $\theta_{3a}$, $\theta_{3b}$ | $V(\theta_{3a})$: 0.94; $V(\theta_{3b})$: 0.23 | $V(\theta_{3b})<V(\theta_{3a})$ |
| In CLC at $\theta_{4a}$, $\theta_{4b}$ | $V(\theta_{4a})$: 0.35; $V(\theta_{4b})$: 0.066 | $V(\theta_{4b})<V(\theta_{4a})$ |
| In CLC at $\theta_{5a}$, $\theta_{5b}$ | $V(\theta_{5a})$: 0.053; $V(\theta_{5b})$: 0.037 | $V(\theta_{5b})<V(\theta_{5a})$ |
| In CLC at $\theta_{6a}$, $\theta_{6b}$ | $V(\theta_{6a})$: 0.022; $V(\theta_{6b})$: 0.006 | $V(\theta_{6b})<V(\theta_{6a})$ |

### 2) Verification of the proposed VFDC

The proposed VFDC strategy is implemented in Cases 4-7 and Case 1 to assess its efficacy. As shown in Fig. 13, without VFDC, the system in Cases 4-7 ultimately stabilizes in the SEP of CLC subsystem. Conversely, with VFDC (Fig. 13(b), Fig. 14(b)), the strategy transforms the $P_{c1}$-$\delta$ curve of the GFMC within the CLC subsystem from sine waves to a straight line with a constant $P_{c1}$ output. This causes the PSL angle to continuously increase until it reaches the switching boundary, $\delta=\delta_k^R$, at which point the system transitions from CLC to CVC. This ensures global stability in CVC. Furthermore, Fig. 14 demonstrates VFDC's effectiveness in enhancing PLL stability. It successfully stabilizes the previously unstable PLL in case 1.

To demonstrate the superiority of the proposed VFDC in stabilizing the GFLC, Fig. 15 (a) compares the PLL frequency response of VFDC against the frequency closed-loop feedback method [27] and the impedance matching method [28] under Case 8. As observed in Fig. 15(a), the method in [27] exhibits high sensitivity to control parameters, where improper PI setting leads to PLL stabilization failure. While the impedance matching method in [28] is effective, it entails complex online impedance estimation and parameter alignment. Conversely, the VFDC eliminates the need for complex tuning or real-time estimation by leveraging local state feedback. Moreover, unlike "PLL freezing" techniques, the VFDC creates a virtually frozen d-axis as depicted in Fig. 9 instead of rigidly locking the loop, allowing the PLL to maintain continuous grid synchronization while ensuring transient stability.

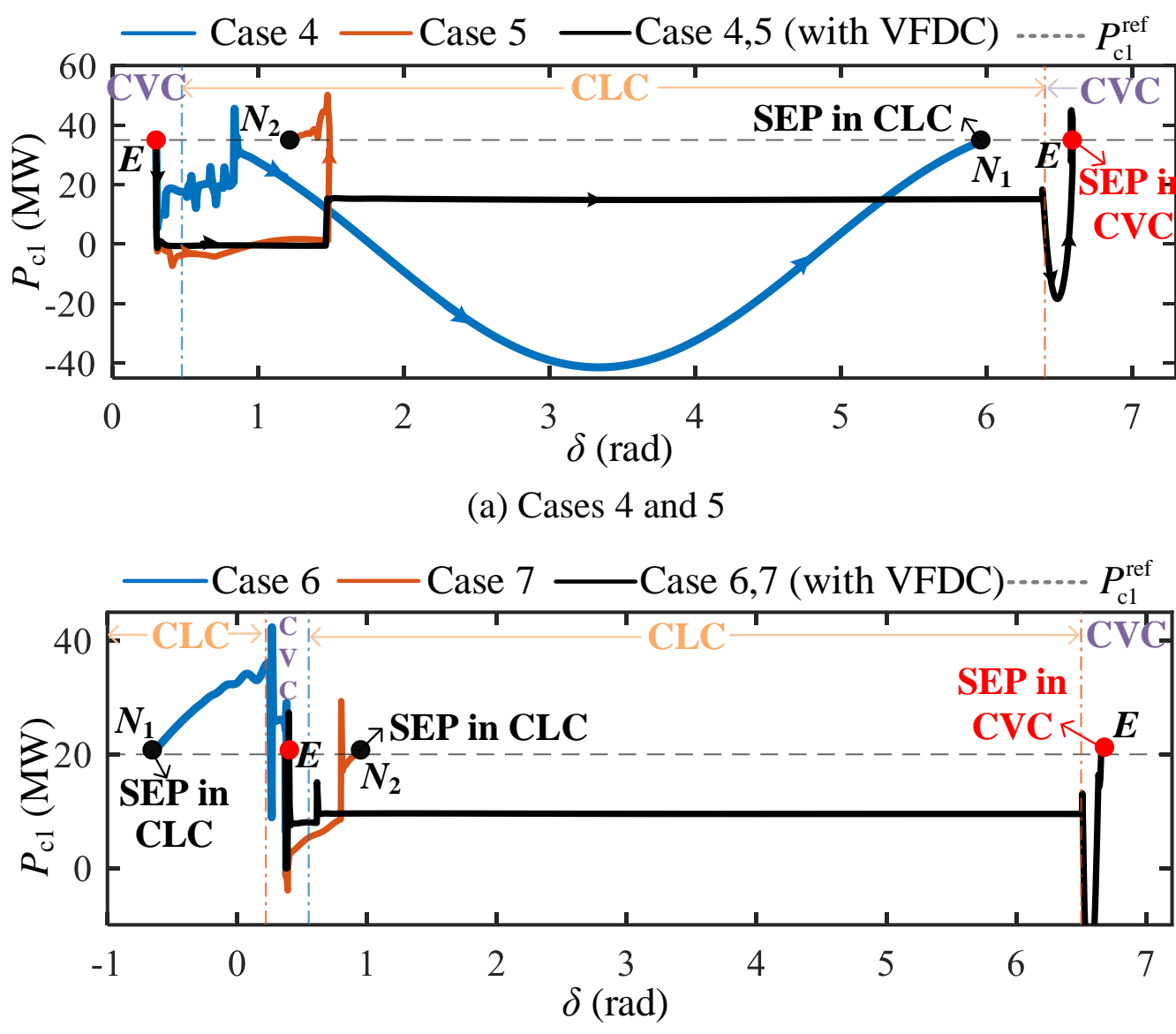

(a) Cases 4 and 5

(b) Cases 6 and 7

**Fig. 13.** GFMC active power versus angle curves with and without VFDC in Cases 4-7.

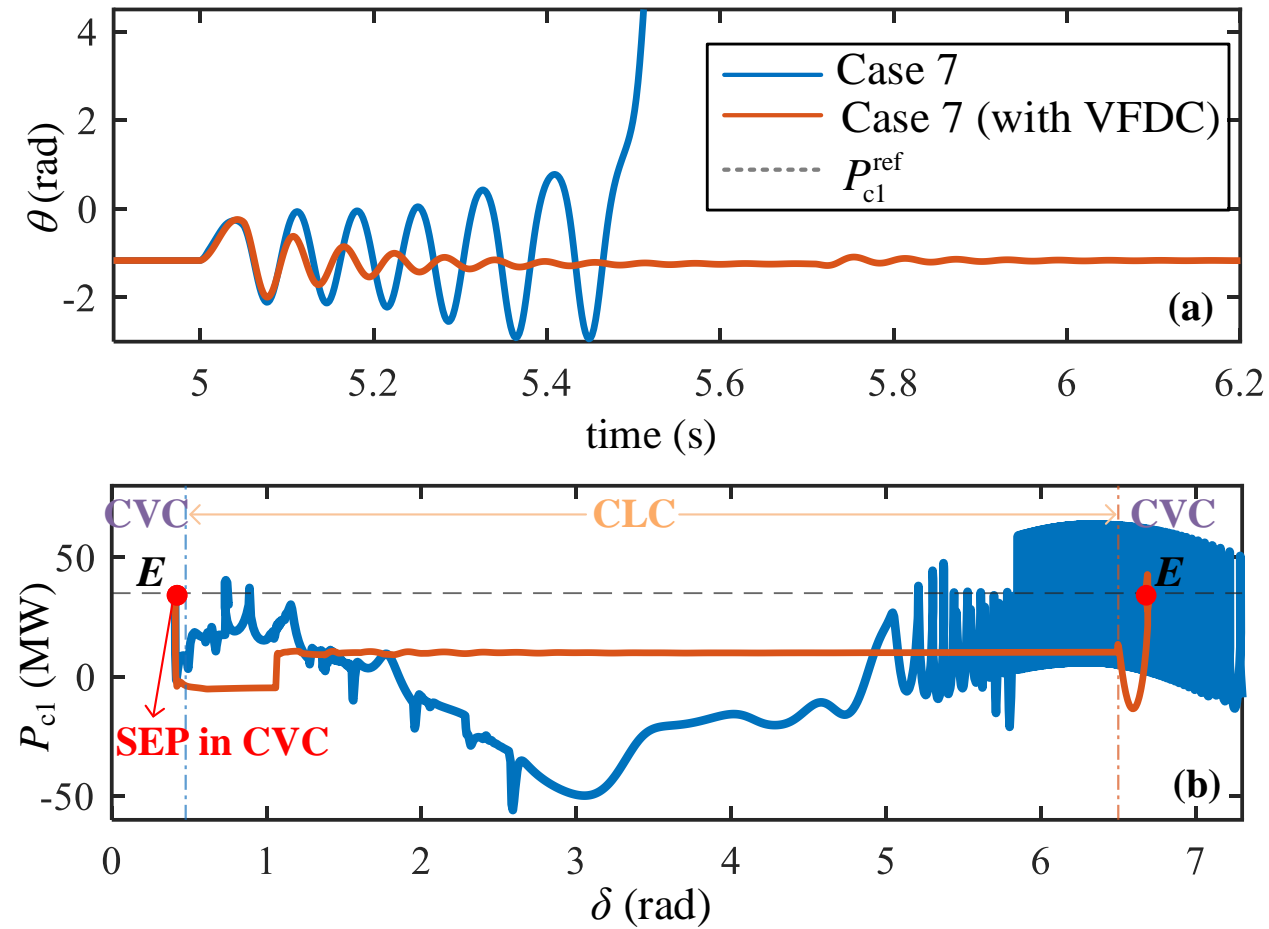

**Fig. 14.** Comparison of results with and without VFDC in Case 1. (a) PLL angle versus time curves; (b) GFMC active power versus angle curves.

Regarding the recovery of PSL-based GFMC from CLC to CVC, Fig. 15(b) presents a comparison between VFDC and the optimal saturation current angle method [18]. This simulation is conducted under Case 8, with the GFLC is disconnected to strictly isolate the GFMC performance. The strategy in [18] relies on pre-calculations based on specific network parameters to ensure SEPs under CLC mode fall within the CVC operation region. However, this dependency compromises robustness against system variations. As illustrated in Fig. 15(b), while both methods work effectively when the post-fault voltage recovers fully (1 p.u.), the method in [18] fails when the voltage recovers to only 0.95 p.u. In the latter scenario, the pre-calculated configuration becomes invalid, causing the GFMC to remain trapped in the CLC mode. In contrast, VFDC successfully restores GFMC to the correct SEP in both scenarios, demonstrating superior adaptability and robustness against operating condition uncertainties.

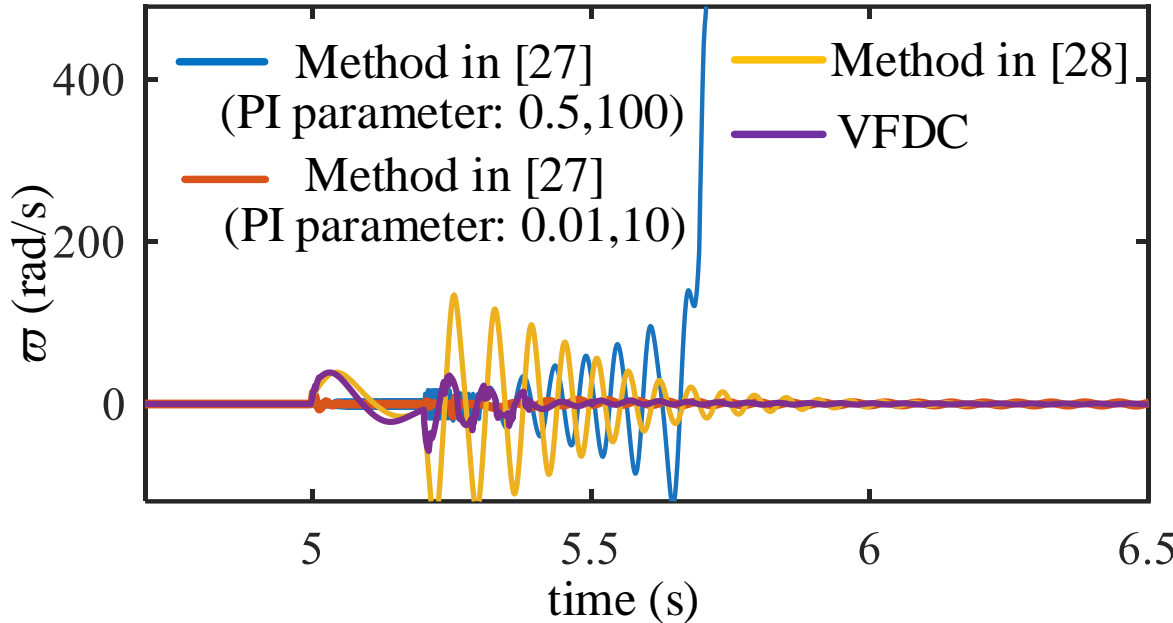

(a) PLL frequency of GFLC under different control strategies.

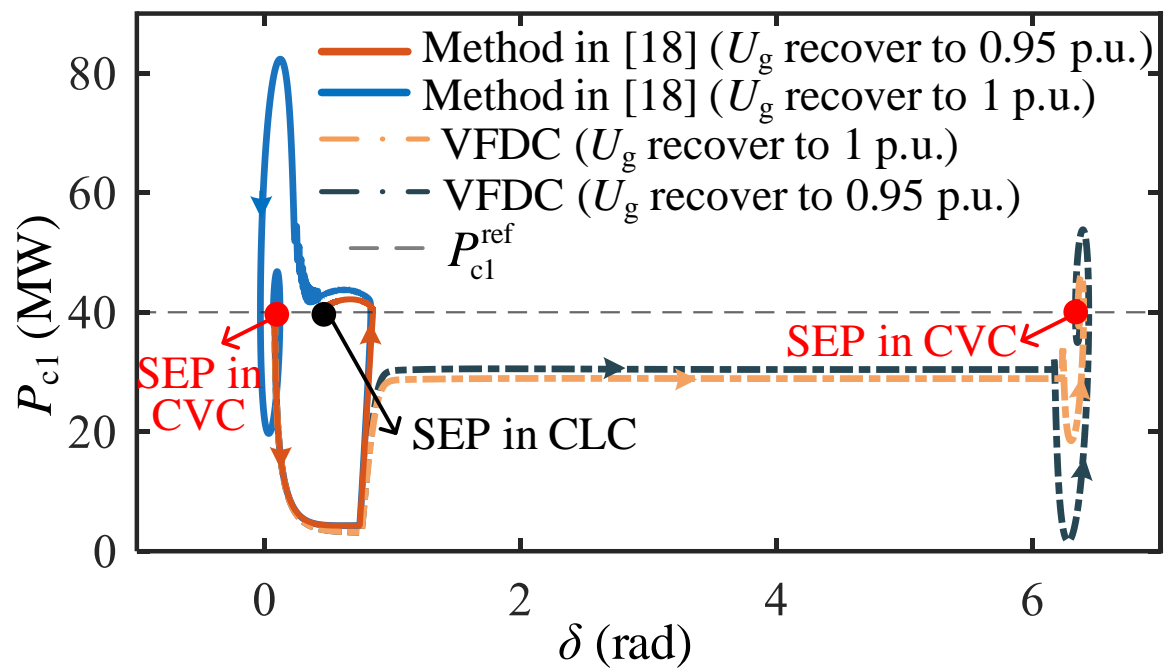

(b) CVC recovery performance of GFMC under different post-fault voltage conditions.

**Fig. 15.** Comparative simulation results of the proposed VFDC strategy and existing methods.

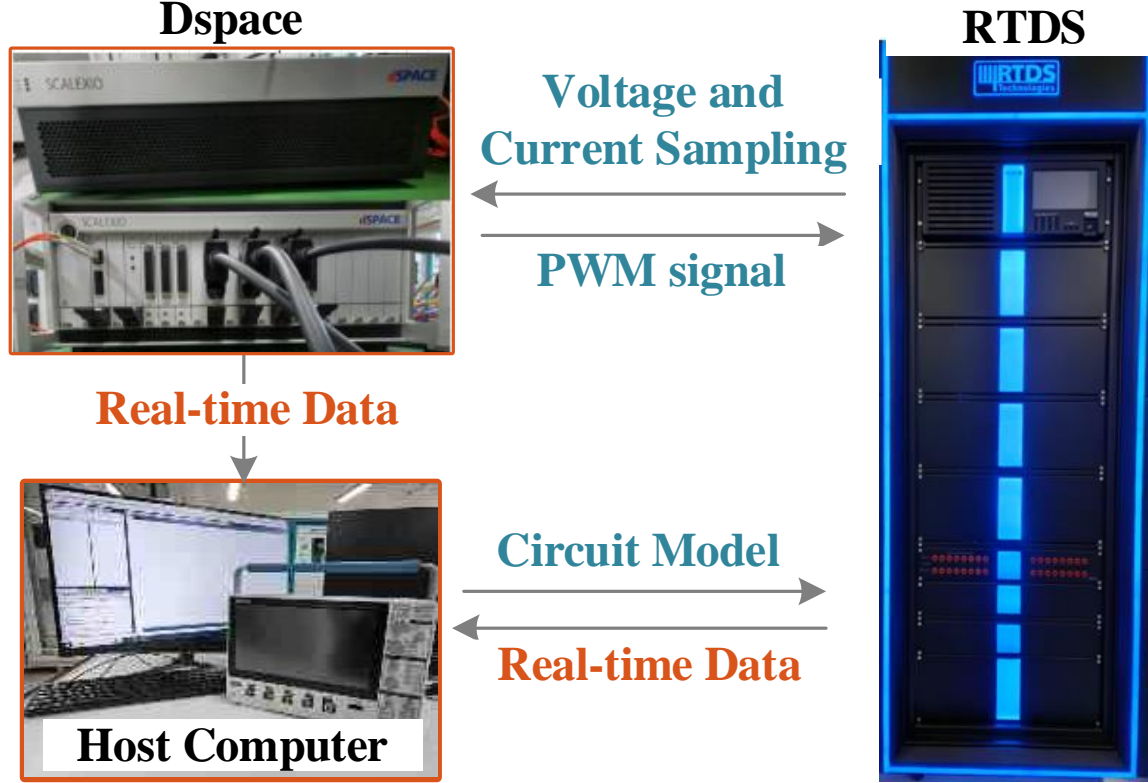

**Fig. 16.** Configuration of the CHIL platform.

*B. Test System 2: Controller Hardware in the Loop Test*

To further validate the proposed VFDC strategy, a controller hardware-in-the-loop (CHIL) platform based on Real-Time Digital Simulator (RTDS) and dSPACE rapid control prototyping system is constructed (Fig. 16). On this platform, the system circuit from Fig. 1(a) is simulated using RTDS, while GFMCs and GFLCs control algorithms are implemented on dSPACE.

Two test cases are examined. In Case 9, GFLCs adopt a single-machine aggregated model, but they are grouped into three clusters with distinct control parameters in Case 10. Relevant system and control parameters are provided in TABLE IV (Appendix D). The fault scenario involves a grid voltage dip to 0.01 p.u. for 200ms.

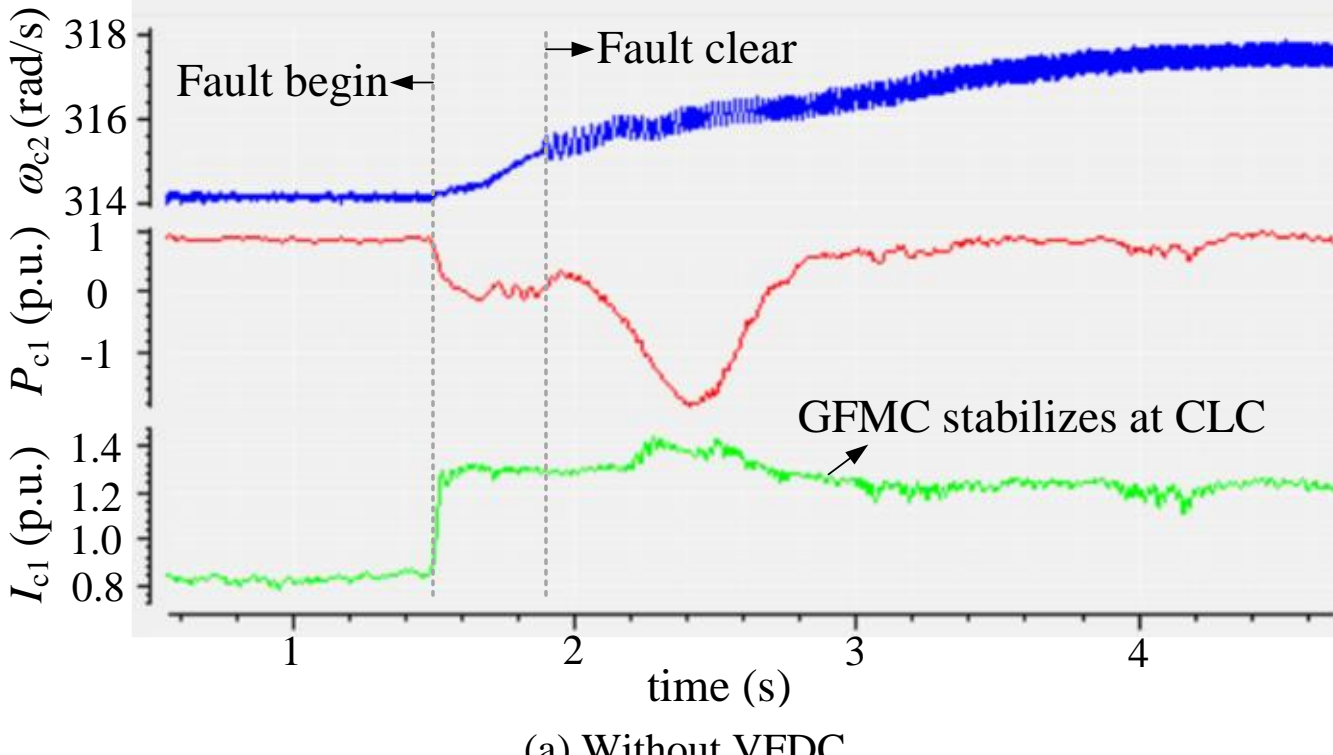

(a) Without VFDC

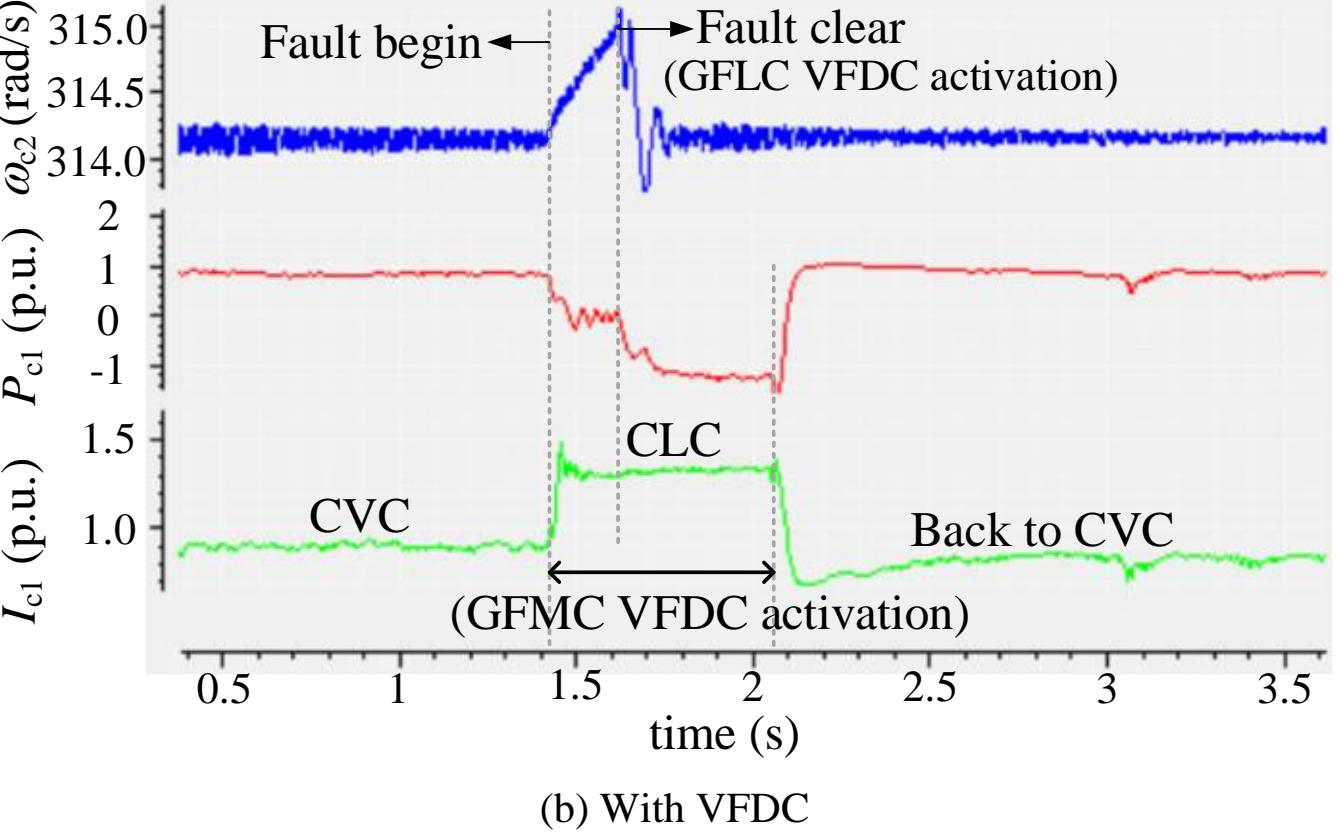

(b) With VFDC

**Fig. 17.** Experiment results of Case 9.

Fig. 17 and Fig. 18 present the CHIL test results for Case 9 and Case 10, respectively. As evident from Fig. 17(a), without the proposed VFDC, the GFLC's PLL became unstable, and the GFMC ultimately stabilized in CLC mode. In contrast, with VFDC applied, the GFLC's VFDC is activated after fault clearance, enabling it to quickly achieve phase synchronization. During the CLC period, the GFMC's VFDC activates to maintain its active power constant, thereby automatically restoring it to CVC mode.

Similarly, for Case 10, Fig. 18 (a) shows that GFLC clusters 1 and 2 lose synchronization when VFDC is not applied. However, with VFDC implemented (Fig. 18(b)), not only do all three GFLC clusters successfully synchronize, but the GFMC also ultimately returned to CVC mode. These results validate the effectiveness of the proposed VFDC strategy in multi-machine systems.

## VI. Discussion

*A. Impact of GFMC Virtual Impedance on GFLC stability*

In grid-following renewable energy systems integrated with GFMCs, the GFMC's primary role is to establish a stable voltage reference for robust phase locking by GFLCs[6]. To achieve this and prevent undesirable dynamic coupling, its voltage and current control loops are designed to operate significantly faster than synchronization loops (PLL and PSL) [22]. Consequently, at the synchronization dynamic timescale, the introduction of a virtual impedance $Z_{\text{virtual}}$ into the GFMC control loop effectively means the GFMC behaves as a voltage source $U_{c1}\angle\delta$ in series with $Z_{\text{virtual}}$. This can be seen as reducing the GFMC's equivalent grid-connection admittance $Y_{c1}$.

Generally, a large virtual impedance is utilized in strong grids to limit short-circuit currents and enhance small-signal stability. In weak-grid conditions, as focused on this paper, GFMCs are expected to provide grid strength support, typically requiring a small $Z_{\text{virtual}}$. If a large $Z_{\text{virtual}}$ is applied in a weak-grid context, making the electrical distance from the GFMC to the PCC non-negligible compared to that from the PCC to the receiving end, its impact on PLL stability is complex. The reduction in $Y_{c1}$ can expand the CVC operating region. However, this comes with a cost. The increased $Z_{\text{virtual}}$ raises $L_v=L_{c1}+(1\text{-}\alpha)\, L_g$ equivalently and thus increases $\theta_s^V$. As a result, both the slope and intercept of $f_v$ increase according to (16), signifying degraded PLL stability. In fact, PLL stability in the CVC subsystem might even fall below that of the CLC subsystem. Moreover, if the increased virtual impedance causes $\theta_s^V$ to exceed $\theta_s^L$, the GFMC's switching behavior could become detrimental. What would normally reduce PLL transient energy might instead increase it.

Therefore, a moderate virtual impedance can expand the CVC operating region, thereby reducing the likelihood of transitioning to CLC mode. However, an excessively large virtual impedance might worsen the sending-end GFLC's synchronous stability in the CVC subsystem. From the perspective of supporting GFLC transient stability, the optimal virtual impedance magnitude warrants further in-depth research.

### *B. Applicability of Conclusions to VSG-Type GFMC*

#### **1) Theoretical validity of PLL stability analysis**

Although the Virtual Synchronous Generator (VSG) introduces second-order angle dynamics due to virtual inertia, this distinction is negligible within the timescale of PLL dynamics. Under CVC operation, both types function as voltage sources. Their CLC behavior is also comparable if adopted same current limiting strategies. Furthermore, in well-designed parallel systems, robust PLL tracking needs PSL or VSG to operate slower than the PLL[13],[22]. This decoupling ensures the GFMC's angle appears approximately "frozen" when analyzing PLL stability. Consequently, the conclusion regarding the switching system remains valid regardless of whether PSL or VSG control is adopted.

#### **2) Applicability of the proposed VFDC strategy**

The VFDC strategy eliminates unintended SEPs in the GFMC's CLC mode, forcing the system trajectory towards the CVC switching boundary to achieve recovery. This directional guidance is sufficient for first-order systems to reach the desired SEP. However, for second-order systems like VSG, merely guiding the trajectory to the CVC region is insufficient if dynamic oscillations persist. VSG synchronization stability demands not only a correct recovery path but also adequate energy dissipation. Therefore, for VSG-GFMC applications, VFDC must be combined with existing stability enhancement measures, such as VSG damping enhancement control[3], to ensure both inherent VSG stability and convergence to the desired equilibrium point.

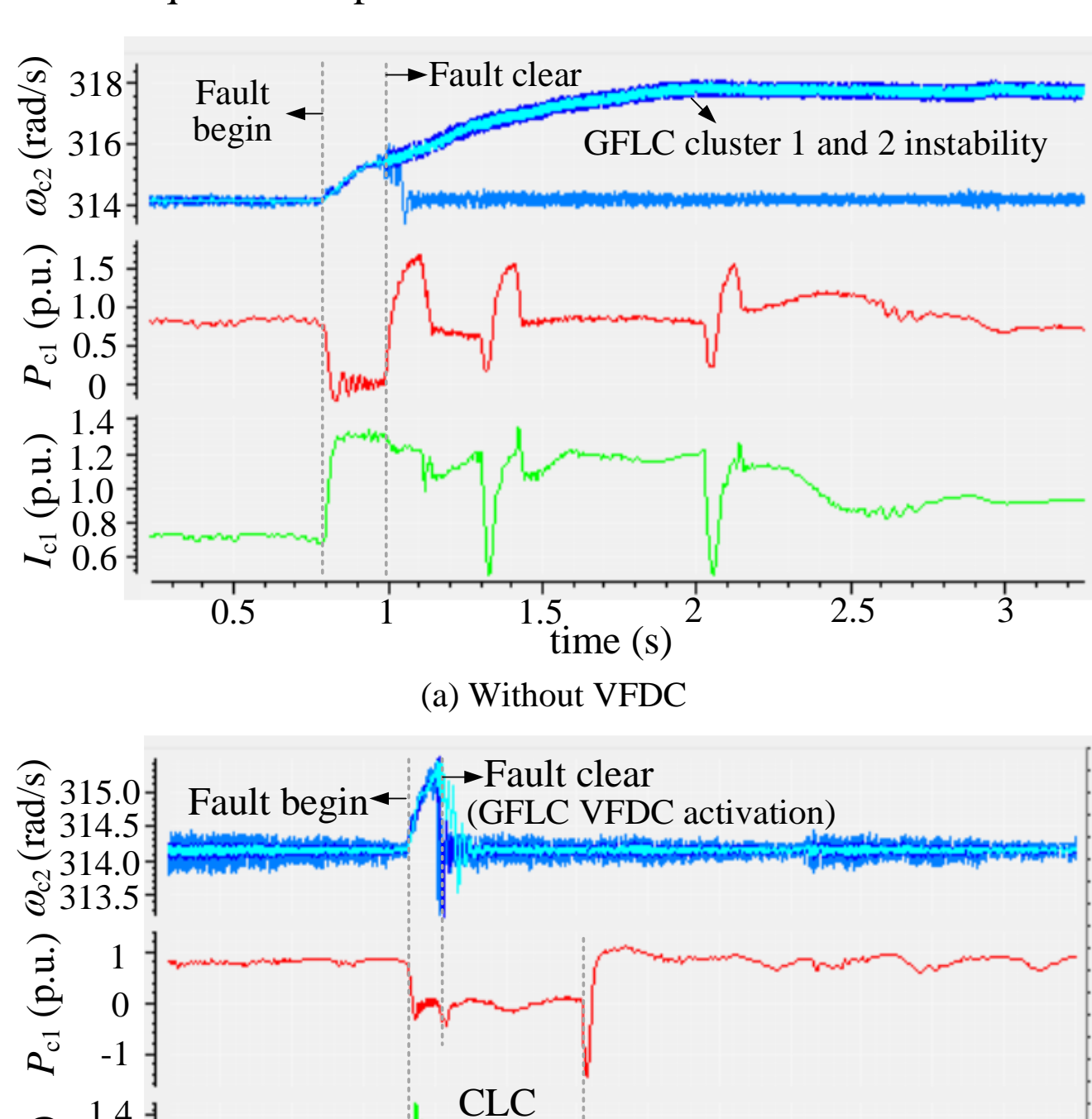

(a) Without VFDC

(b) With VFDC

**Fig. 18.** Experiment results of Case 10.

## VII. CONCLUSION

Upon integration of GFMCs, their current saturation characteristics transform the GFLC-dominated power system into a state-dependent switching system. There are two distinct switching modes exist: "GFMC angle-dependent switching" at the PSL dynamic timescale and "GFLC angle-dependent switching" at the PLL dynamic timescale. Under these dynamics, the system faces two primary stability challenges: the failure of GFMCs to recover to CVC mode post-fault, and the PLL transient instability constrained by the CLC boundary. To address these issues, this paper proposes a novel control concept: virtual fixed d-axis control (VFDC). Relying solely on local state feedback, VFDC adaptively facilitates PLL convergence and ensures the correct restoration of the GFMC to CVC mode. Comparative analysis confirms that the proposed strategy achieves robust and adaptive stabilization with a simple structure compared to existing methods

## APPENDIX

### *A. Expressions for Variables in* (6)

$$P_{\text{Ec2}}^{\text{V}} = \alpha U_{\text{c1}} \sin(\theta-\delta) + (1-\alpha) U_{\text{g}} \sin\theta\,, \tag{26}$$

$$P_{\text{Mc2}}^{\text{V}} = \omega_{\text{g}} L_{\text{v}} i_{\text{c2}}^{\text{d}*}\,, \tag{27}$$

$$T_{\text{c2}}^{\text{V}} = \left[1 - k_{2\text{p}}(1-\alpha) L_{\text{v}} i_{\text{c2}}^{\text{d}*}\right] / k_{2\text{I}}\,, \tag{28}$$

$$D_{\text{c2}}^{\text{V}} = k_{2\text{p}} / k_{2\text{I}} \left[\alpha U_{\text{c1}} \cos(\theta-\delta) + (1-\alpha) U_{\text{g}} \cos\theta\right] - L_{\text{v}} i_{\text{c2}}^{\text{d}*}\,, \tag{29}$$

$$P_{\mathrm{Ec2}}^{\mathrm{L}} = U_{\mathrm{g}} \sin\theta \,, \tag{30}$$

$$P_{\mathrm{Mc2}}^{\mathrm{L}} = \omega_{\mathrm{g}} L_{\mathrm{g}} i_{\mathrm{c1}}^{\mathrm{d}*} + \omega_{\mathrm{g}} (L_{\mathrm{g}} + L_{\mathrm{c2}}) i_{\mathrm{c2}}^{\mathrm{q}*} \,, \tag{31}$$

$$D_{\mathrm{c2}}^{\mathrm{L}} = k_{\mathrm{2p}} / k_{\mathrm{2I}} \left[ U_{\mathrm{g}} \cos\theta + \omega_{\mathrm{g}} L_{\mathrm{g}} i_{\mathrm{c1}}^{\mathrm{q}*} - (L_{\mathrm{g}} + L_{\mathrm{c2}}) i_{\mathrm{c2}}^{\mathrm{d}*} \right], \tag{32}$$

$$D_{\mathrm{c2}}^{\mathrm{L}} = k_{\mathrm{2p}} / k_{\mathrm{2I}} \left[ U_{\mathrm{g}} \cos\theta + \omega_{\mathrm{g}} L_{\mathrm{g}} i_{\mathrm{c1}}^{\mathrm{q}*} - (L_{\mathrm{g}} + L_{\mathrm{c2}}) i_{\mathrm{c2}}^{\mathrm{d}*} \right], \tag{33}$$

### *B. Proof of $\theta_1 < \theta_s^V < \theta_2$*

From (11), it is clear that the PLL's SEP in CVC always exists, regardless of the value of $\delta$. Fig. 19(a) illustrates the vector relationships at the switching instant within the CVC subsystem. As shown, for a closed loop to be formed by the vectors, $\theta_{\mathrm{s}}^{\mathrm{V}}$ must necessarily fall within the range $\theta_1 < \theta_{\mathrm{s}}^{\mathrm{V}} < \theta_2$. Otherwise, eq. (9) would be unsolvable.

### *C. Proof: $\theta_s^V \leq \theta_s^L$ at Switching Moments*

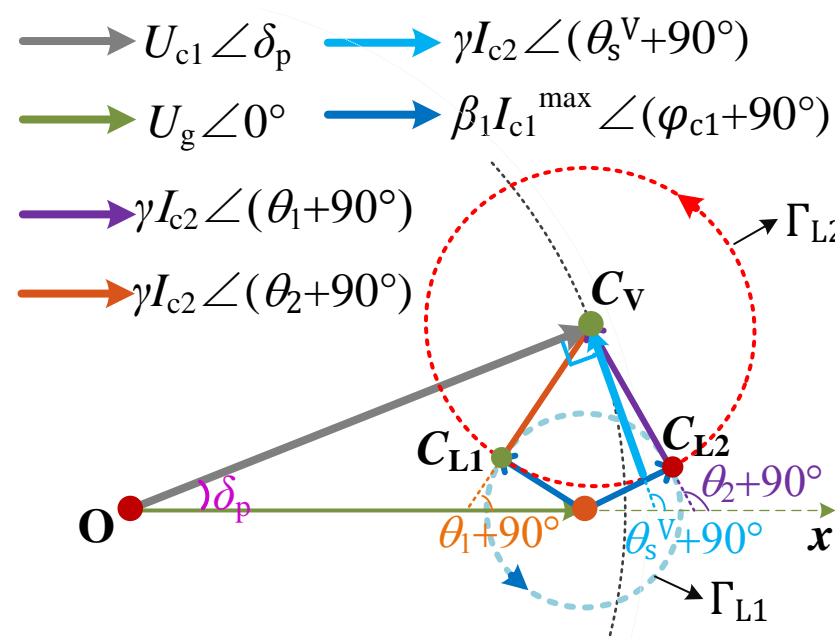

(a) The relation between switching points and the SEP of PLL

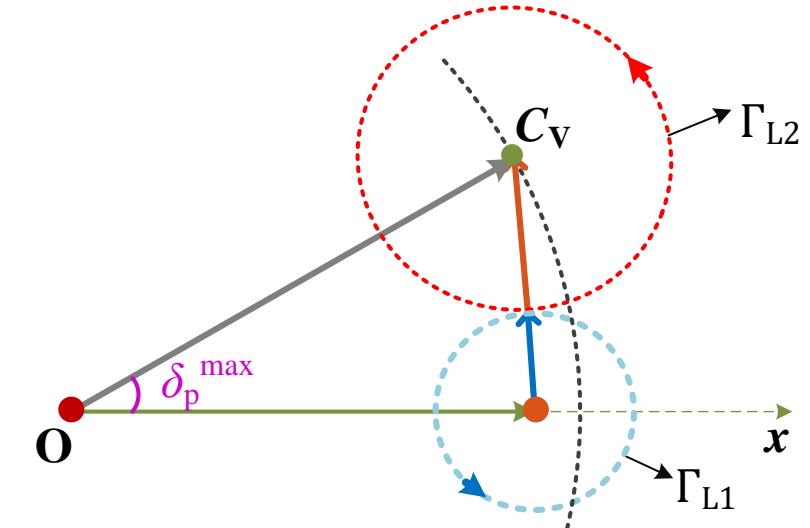

(b) Vector relationships when $\delta_p$ reaches its maximum value

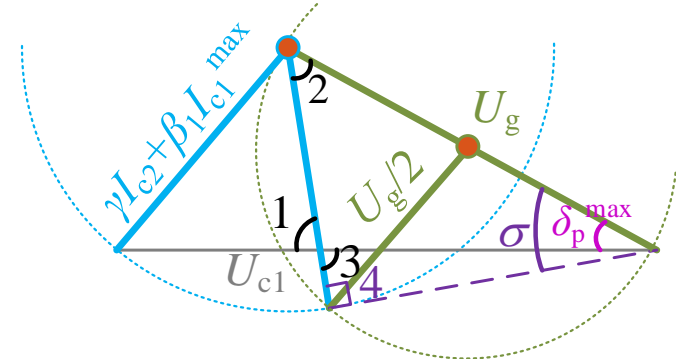

(c) Relationships between $\delta_p^{max}$ and $\sigma$

**Fig. 19.** Vector relationships in the CVC subsystem at switching instant.

For a conservative analysis, $i_{\mathrm{c1}}^{\mathrm{d}_2}$ can be approximated by its maximum value, $I_{\mathrm{c1}}^{\max}$, which represents the most critical operating condition for PLL stability. In (11) and (12), given that $L_{c1}$ is negligible compared to $L_g$, $\theta_{\mathrm{s}}^{\mathrm{V}}$-$\theta_{\mathrm{s}}^{\mathrm{L}}$ is expressed as:

$$\theta_{\mathrm{s}}^{\mathrm{V}} - \theta_{\mathrm{s}}^{\mathrm{L}} \approx \delta_{\mathrm{p}} - \sigma = \delta_{\mathrm{p}} - \arcsin \frac{\beta_1 I_{\mathrm{c1}}^{\max} + \gamma i_{\mathrm{c2}}^{\mathrm{d}_2}}{U_{\mathrm{g}}} \,. \tag{34}$$

From Fig. 19(b), $\delta_p$ reaches its maximum, $\delta_{\mathrm{p}}^{\max}$, if $\beta_1 I_{\mathrm{c1}}^{\max} \angle (\varphi_{c1}+90°)$ is collinear with $\gamma I_{c2} \angle (\varphi_{c2}+90°)$. Let $\sigma$ be the angle in the second term on the right side of (34). The geometric relationship between $\sigma$ and $\delta_{\mathrm{p}}^{\max}$ is shown in Fig. 19(c). It can be proven that $\delta_{\mathrm{p}}^{\max} \leq \sigma$ (Using proof by contradiction: if $\sigma < \delta_{\mathrm{p}}^{\max}$, then $\angle 3 = \angle 1 = \angle 2 + \delta_{\mathrm{p}}^{\max} > 90°$, which contradicts the fact that $\angle 3 < \angle 4 = 90°$. Thus, $\delta_{\mathrm{p}}^{\max} \leq \sigma$). Therefore, as per (34), $\theta_{\mathrm{s}}^{\mathrm{V}} \leq \theta_{\mathrm{s}}^{\mathrm{L}}$.

### *D. Parameters of the Test Systems*

TABLE III
PARAMETERS OF TEST SYSTEM 1

| Item | Parameter Name | Parameter Value |
| --- | --- | --- |
| GFLC | PI parameters of outer power loop | 0.005,20 |
| | PI coefficient of the PLL | 0.07 or 0.15,100 |
| | PI parameters of inner current loop | 70,125 |
| | Filter inductor, resistor | 0.065,0.001 p.u. |
| GFMC | Droop coefficient of PSL | 0.04 |
| | PI parameters of inner voltage loop | 0.015,0.2 |
| | PI parameters of inner current loop | 60,120 |
| | Filter inductor, capacitor, resistor | 0.15,50.59,0.005 p.u. |
| Branch | $1/Y_{c1}$, $1/Y_{c2}$, $1/Y_g$ | 0.05, 0.03, 0.58 p.u. |
| Grid | Voltage level | 220 kV |
| | Base frequency and capacity | 50 Hz, 200 MVA |

TABLE IV
PARAMETERS OF TEST SYSTEM 2

| Item | Case | Parameter Name | Parameter Value |
| --- | --- | --- | --- |
| GFLC | 8 | PI coefficient of the PLL | 0.25, 100 |
| | | PI parameters of current loop | 80, 500 |
| | | Filter inductor | 7e-3 H |
| | | Output power (7 units) | 60.2 MW |
| | 9 | PI coefficient of the PLL | Cluster 1: 0.25, 100; Cluster 2: 0.15, 50; Cluster 3: 0.6, 100. |
| | | PI parameters of current loop | 150, 100 |
| | | Filter inductor, resistor | 7e-3 H |
| | | Output power | Cluster 1: 27.9 MW; Cluster 2: 16.8 MW; Cluster 3: 8.4 MW. |
| PSL-GFMC | | Droop coefficient of PSL | 0.02 |
| | | PI parameters of voltage loop | 1, 1.5 |
| | | PI parameters of current loop | 150, 150 |
| | | Filter inductor, resistor | 5e-3 H, 7e-5 Ω |
| | | Rated capacity (2 units) | 12 MVA |
| | | Current saturation angle | 0 rad |
| Branch | | $1/Y_{c1}$, $1/Y_{c2}$, $1/Y_g$ | 0.26, 0.05, 0.4 p.u. |
| Grid | | Voltage level | 220 kV |
| | | Base frequency and capacity | 50 Hz, 200 MVA |

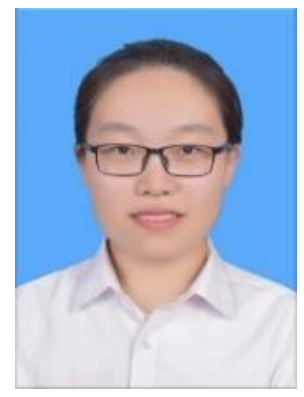

**Bingfang Li** (Member, IEEE), received the B.S. degree from North China Electric Power University, Baoding, China, in 2022, and is currently working toward the Ph.D. degree with Xi’an Jiaotong University. Her main fields of interest include Power system stability analysis and control.

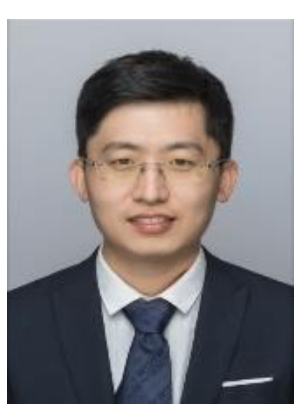

**Songhao Yang** (Senior Member, IEEE) was born in Shandong, China, in 1989. He received the B.S. and Ph.D. degrees in electrical engineering from Xi’an Jiaotong University, Xi’an, China, in 2012 and 2019, respec-tively, and the Ph.D. degree in electrical and electronic engineering from Tokushima University, Tokushima, Japan, in 2019. He is currently an Associate Professor with Xi’an Jiaotong University. His research focuses on power system stability analysis and control.

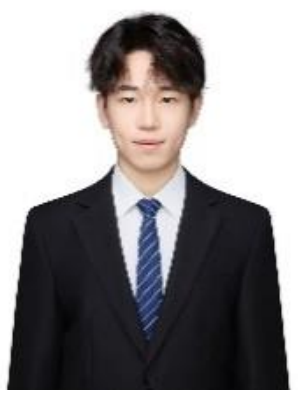

**Pu Cheng** (Graduate Student Member, IEEE), received the B.S. degree from Xi’an Jiaotong University, Xi’an, China, in 2024, and is currently working toward the Ph.D. degree with Xi’an Jiaotong University. His main fields of interest include Power system voltage stability analysis.

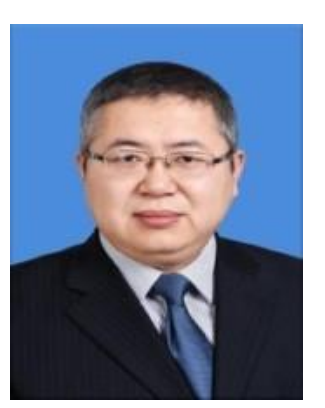

**Zhiguo Hao** (Senior Member, IEEE), was born in Ordos, China, in 1976. He received the B.Sc. and Ph.D. degrees in electrical engineering from Xi’an Jiaotong University, Xi’an, China, in 1998 and 2007, respectively. He is currently a Professor with the Electrical Engineering Department, Xi’an Jiaotong University. His research focuses on power system protection and control.